\documentclass[11pt,a4paper]{article}
\usepackage{graphicx}
\usepackage{jcappub}
\usepackage{lscape}

\usepackage{amsmath, amssymb}
\usepackage[T1]{fontenc}
\usepackage[utf8]{inputenc}
\usepackage{lmodern}
\usepackage[english]{babel}
\usepackage{float}

\usepackage[normalem]{ulem}
\usepackage{subfigure}
\usepackage{xcolor}
\usepackage{amsmath}
\usepackage{amssymb}
\usepackage{bm}

 \makeatletter
\def\alt{\mathrel{\mathpalette\gl@align<}}
\def\agt{\mathrel{\mathpalette\gl@align>}}
\def\gl@align#1#2{ \lower.6ex\vbox{\baselineskip\z@skip\lineskip\z@
\ialign{ $\m@th#1\hfil##\hfil$\crcr#2\crcr\sim\crcr }} } \makeatother

\title{Higher multipoles of the galaxy bispectrum in redshift space
}

\author{Yue Nan,${}^a$ Kazuhiro Yamamoto,${}^a$ Chiaki Hikage${}^b$
}

\affiliation{
  ${}^a$Department of Physics, Graduate School of Physical Sciences, Hiroshima University,
  Higashi-Hiroshima, Kagamiyama 1-3-1, 739-8526, Japan\\
  ${}^b$Kavli Institute for the Physics and Mathematics of the Universe (Kavli IPMU, WPI), \\
The University of Tokyo, 5-1-5 Kashiwanoha, Kashiwa, Chiba, 277-8583, Japan}


\abstract{
As a generalization of our previous work [Phys. Rev. D 95 043528 (2017)], in which an analytic model
for the galaxy bispectrum in redshift space was developed on the basis of the
halo approach, we here investigate its higher multipoles that have not been known so far.
The redshift-space bispectrum includes the two variables $\omega$ and $\phi$
for the line-of-sight direction, and the higher multipole bispectra are defined
by the coefficients in the expansion of the redshift-space bispectrum
using the spherical harmonics.
We find 6 new nonvanishing components out of $25$ total components up to $\ell=4$,
in addition to 3 components discussed in the previous work (monopole,
quadruple, and hexadecapole of $m=0$).
The characteristic behaviors of the new nonvanishing multipoles are compared with the results of
galaxy mock catalogs that match the halo occupation distribution of the Sloan Digital Sky Survey Baryonic Oscillation Spectroscopic Survey low-redshift sample.
We find that the multipoles with $m\ne 0$ are also sensitive to redshift-space distortion (RSD) as well as those with $m=0$ and thus are key ingredients in the RSD analysis using the galaxy bispectrum.
Analytic approximation formulas for these nonzero components are also presented;
these are useful for understanding the characteristic behaviors.
}

\begin{document}
\maketitle

\section{Introduction}

In statistical analysis of the large-scale structure of galaxies, the basic
quantities are the two-point correlation function and the power spectrum
which are related by the Fourier transformation.
If the fluctuations are statistically isotropic and Gaussian, the monopole
power spectrum should be enough to characterize the statistical properties.
However, non-Gaussian properties might have been imprinted in the initial
conditions of primordial fluctuations \cite{Planckresults}. Furthermore,
in the course of evolution of cosmic structure formation, the non-Gaussian
properties are generated in the density perturbation and in the galaxy
distributions owing to the nonlinearity of gravitational clustering and
structure formation (e.g., \cite{BS1,BS2,BS3,IBispec1}).
The three-point correlation function in the configuration space and its Fourier
transformation, i.e., the bispectrum, is the lowest order statistical quantity
used to characterize these non-Gaussian properties (see, e.g.,
\cite{Scoccimarro2000} for a review)
The first measurements of the three-point correlation function were reported in Refs.
\cite{PeeblesGroth,GrothPeebles}, thereafter many works have been carried out
\cite{Kayo2004,Gaztanaga2005,Nichol2006,Kulkarni2007,Gaztanaga2009,McBride2011a,McBride2011b,Marin2011,Marin2013,Guo2014,Guo2015,GagranSamushia2017,PearsonSamushia2017,SlepianEisenstein2015,SlepianEisenstein2016,Slepianelal2017a,Slepianelal2017b,SlepianEisenstein2017}.
On the other hand, the measurement of the bispectrum was carried out for the first time
by Fry and Seldner \cite{FrySeldner}, and it was advanced as in the
Refs.~\cite{Scoccimarro2001M,Feldman2001,Verde2002,HGilMarin1,HGilMarin2,HST,Sugiyama}.

In the analysis of the higher order statistics with galaxy catalogs of redshift surveys,
peculiar velocities of galaxies break the assumption of statistical isotropy
in the distribution in redshift space through redshift-space distortion.
On large scales, the linear velocity field induces a linear redshift-space
distortion \cite{Kaiser1987,Hamilton};
however, on small scales, the random velocity of galaxies make a significant
contribution to the distribution in redshift space, which is called the Finger
of God (FoG) effect \cite{Jackson,PeacockDodds}.
The redshift-space distortion generates the additional non-Gaussianity
in the galaxy distribution in redshift space \cite{Scoccimarro1999}.
Especially, the FoG effect is the cause of the non-Gaussianity on small scales, reflecting
the nonlinear evolution of the cosmic structure formation as well.
Recently, higher multipoles which characterize the redshift space-distortions
in the three-point correlation function and the bispectrum are discussed
\cite{Scoccimarro2015,YNH,SlepianEisenstein2017,HST,Sugiyama}.

In the present paper, we focus on the galaxy bispectrum in redshift space.
A precise theoretical model
is necessary to obtain cosmological information beyond the two point statistics
from the galaxy bispectrum
\cite{Shirata,Koyama,Barreira,Emilio1,Emilio2,Takushima1,Takushima2,Munshi,Yamauchi,Hirano}
Recently, various general relativistic effects and the wide angle effect
in the bispectrum are also discussed
\cite{Raccanelli,Umeh,Jolicoeur1,Jolicoeur2,Dio1,Dio2,Bertacca}.
However, the bispectrum is quite complicated even in the simplest case
within the general relativity. Therefore, an analytic model that
reproduces their behaviors well is quite useful.
The halo approach is useful to find such a theoretical model that is applicable from large
to small scales \cite{White,Seljak,Scoccimarro2001,CooraySheth2002}.
The theoretical framework, based on the assumption that all the dark matter and
galaxies are associated with virialized dark matter halos, is characterized by
the halo density profile $\rho(r)$, the halo mass function $dn\slash dM$, and the halo's
correlation, where $\rho(r)$ represents the density of each halo and $dn\slash dM$ represents
the number density of halos with mass $M$.
In Ref. \cite{HY}, the authors developed a theoretical model to explain the
multipole power spectra in redshift space of the Sloan Digital Sky Survey (SDSS)
luminous red galaxies (LRGs) on the basis of
the halo approach, in which the halo occupation distribution (HOD) of the central galaxy
and satellite galaxies plays an important role. The theoretical model well reproduces
the results of the observational data.  In our previous paper \cite{YNH}, the theoretical
approach was applied to the model for the galaxy bispectrum in redshift space.
We demonstrated that the theoretical model reproduces the behaviors of the bispectrum
of mock galaxy catalogs of the low-redshift (LOWZ) galaxy sample of the SDSS III
Baryon Oscillation Spectroscopic Survey (BOSS) survey \cite{Parejko}.
An advantage of the theoretical approach is that analytic approximate expressions
for the bispectrum
can be obtained, which is useful for understanding how the bispectrum depends
on the parameters qualitatively.

As a generalization of the previous work \cite{YNH}, we investigate higher
multipoles of the bispectrum.
We find the nonvanishing higher multipole components of the bispectrum, which have not been
known so far. Such components of the bispectrum will be useful for characterizing the
unique non-Gaussian properties of the galaxy distribution in redshift space.
This paper is organized as follows. In Section 2, we introduce the multipoles of the
bispectrum as coefficients in the multipole expansion with respect to the spherical harmonics
$Y_{\ell,m}(\omega,\phi)$, where $\omega$ and $\phi$ are the parameters for the line-
of-sight direction. Previous works only investigated the components of $m=0$
\cite{Scoccimarro1999,Scoccimarro2015,YNH}. We find six new components of real functions
up to $\ell=4$.
In Section 3, the behaviors of the new multipoles of the bispectrum are demonstrated
by adopting the HOD of the SDSS III BOSS LOWZ sample.
In Section 4, the properties of the multipoles of the bispectrum are investigated
in an analytic way. Section 5 is devoted to a summary and conclusions.
Appendix A lists the expression for the spherical harmonics. Appendix B lists the analytic formulas for the multipoles of the bispectrum
in redshift space.
In the present paper, we adopt a spatially flat cold dark matter (CDM) cosmology
with a cosmological constant $\Lambda$ adopting the parameters
$\Omega_b=0.046$, $\Omega_m=0.273$, $n_s=0.963$, $h=0.704$, $\tau=0.089$, and
$\sigma_8=0.809$.

\section{Basis for bispectrum in redshift space}
\subsection{Halo model}

We first introduce the bispectrum in redshift space in the halo approach.
The halo approach is quite useful for characterizing the distributions of dark matter
as well as the distributions of galaxies, from large scales to smaller scales, where nonlinearity
plays an important role \cite{White,Seljak,Scoccimarro2001,CooraySheth2002,smith2008,HY}.
In the present paper, we follow the theoretical model developed in Ref.~\cite{YNH},
where an analytic expression was presented for the bispectrum in which the halo approach was applied,
with the HOD description of central galaxies and satellite galaxies.
By adopting the model of \cite{YNH}, a generalized model will be developed.

As addressed previously, the basic quantities used in the halo approach are the halo density profile
$\rho(r)$ characterizing the matter distribution within halos and the halo mass function
$dn\slash dM$ describing the distribution of halos themselves.
In addition, random motions of galaxies within halos, as an embodiment of nonlinearity on small scales,
are characterized by assuming an uncorrelated one-dimensional velocity dispersion yielding a Gaussian distribution.
The HOD offers a method for linking statistical quantities of galaxies to that of halos in the halo approach.

For the halo density profile $\rho(r)$, assuming the truncated
Navarro--Frenk--White (NFW) density profile \cite{NFWprofile} of dark matter, we can write
\begin{eqnarray}
\rho(r)=\left\{
\begin{array}{ll}
\displaystyle{\rho_s\over r/r_s(1+r/r_s)^2}  & (r<r_{\rm vir}),\\
0 & (r>r_{\rm vir}),
\end{array}
\right.
\end{eqnarray}
where $\rho_s$ and $r_s$ are the parameters representing the characteristic density and
the characteristic scale, and $r_{\rm vir}$ is the virial radius, which determines
the virial mass of a halo by $M_{\rm vir}=4\pi\int_0^{r_{\rm vir}} dr r^2\rho(r)
=4\pi r_{\rm vir}^3 \Delta_{\rm vir} \bar \rho_m (z)/3$, where $\bar \rho_m(z)$ is
the mean matter density and we adopt the value $\Delta_{\rm vir}=265$ at redshift $z=0.3$.
Because our interest is focused on the quantity in Fourier space, we denote the Fourier transform of $\rho(r)$ by
\begin{eqnarray}
\tilde u_{\rm NFW}(k;M)=&\displaystyle{\int_{r\leq r_{\rm vir}}d^3x \rho(r) e^{-i{\bm k}\cdot{\bm x}}\over
\int_{r\leq r_{\rm vir}}d^3x \rho(r)}.
\end{eqnarray}

For the distribution of halos, we adopt the fitting formula in
Refs.~\cite{ShethTormen1999,ShethTormen2002,Parkinson2007}
for the halo mass function in the form
\begin{eqnarray}
 M{dn\over dM}={\bar \rho_m \over M}{d\ln \sigma_R^{-1}\over d\ln M} f(\sigma_R)
\end{eqnarray}
with
\begin{eqnarray}
 f(\sigma_R)=0.322\sqrt{2\times0.707\over \pi}\left[
1+\left({1\over 0.707\nu^2}\right)^{0.3}\right]{\nu}\exp
\left(-{0.707\nu^2\over 2}\right)
\label{massfunctionmodel}
\end{eqnarray}
and  $\nu = \delta_c/\sigma_R$, where $\sigma_R$ is the root-mean-square fluctuation in spheres
containing mass $M$ at the initial time, extrapolated to redshift $z$ using linear theory, and
$\delta_c(\simeq1.686)$ is the critical value of the initial overdensity
that is required for gravitational collapse.

We assume that the distribution of satellite galaxies follows the NFW profile and
that the satellite galaxies have internal random velocities following a Gaussian
distribution specified by the one-dimensional velocity dispersion
\cite{HY,Kanemaru,ReidSpergel,LokasMamon},
\begin{eqnarray}
\sigma_{v,{\rm off}}(M)=\left({GM\over2r_{\rm vir}}\right)^{1/2}.
\label{sigmavoff}
\end{eqnarray}
These random motions cause the FoG effect,
which changes the distribution of satellite galaxies in redshift space.
If satellite motions in a halo are uncorrelated with each other, then the Fourier transform of the distribution of the satellite
galaxies in redshift space yields
\begin{eqnarray}
\label{eq:psoff}
\widetilde u(\bm k,M)&=&\tilde{u}_{\rm NFW}(k;M)\exp\left[-\frac{\sigma_{v,{\rm off}}^2(M)k^2\mu^2}{2a^2H^2(z)}\right],
\label{eq:psoff2}
\end{eqnarray}
where $H(z)$ is the Hubble parameter at the redshift $z$.

\begin{table}[b]
\begin{center}
\begin{tabular}{cccc}
\hline\hline
$~~~~~~~~~M_{\rm min}$~~~~ & ~~~~$1.5\times 10^{13}h^{-1}\;M_\odot$~~~~~~~~~ \\
$\sigma_{\log M}$ & 0.45  \\
$M_{\rm  cut}$ &  $1.4\times 10^{13}h^{-1}\;M_\odot$ \\
$M_1$ & $1.3\times10^{14}h^{-1}\;M_\odot$  \\
$\alpha$ & $1.38$ \\
\hline\hline
\end{tabular}
\caption{HOD parameters for the LOWZ sample \cite{Parejko}.}
\end{center}
\label{tab:lrg_halo}
\end{table}

To link the distribution of satellite galaxies in redshift space to
that of halos, we introduce the halo occupation distribution $N_{\rm HOD}(M)$, which describes the average
occupation number of galaxies inside a halo with mass $M$. We adopt the following fitting formula for
central galaxies and satellite galaxies \cite{Zheng2005}:
\begin{eqnarray}
  &&N_{\rm HOD}(M)=\langle N_{\rm c}\rangle(1+\langle N_{\rm s}\rangle),
\label{NHODM}
  \\
&&\langle N_{\rm c}\rangle =\frac{1}{2}\left[1+{\rm erf}\left(\frac{\log_{10}(M)-\log_{10}
      (M_{\rm min})}{\sigma_{\log M}}\right)\right],
  \label{defNc}
  \\
&&\langle N_{\rm s}\rangle =
  \left(\frac{M-M_{\rm cut}}{M_1}\right)^{\alpha},
\label{eq:HOD}
\end{eqnarray}
where ${\rm erf}(x)$ is the error function.
For specific values, the HOD parameters are listed in Table I for the
SDSS-III BOSS LOWZ catalog \cite{Parejko}.

\begin{figure}[b]
\begin{center}
\includegraphics[width=80mm]{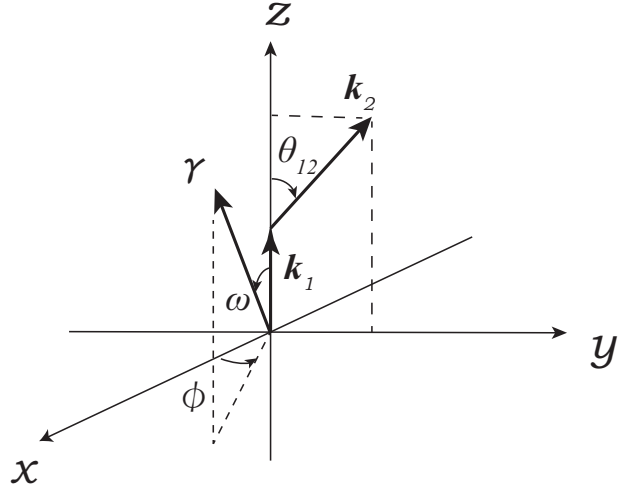}
\caption{Definition of variables for the bispectrum.}
\label{fig:configuration}
\end{center}
\end{figure}

\subsection{Bispectrum in redshift space}
If we denote the galaxy number density contrast by $\delta(t,\bm k)$, the bispectrum
$B_g(t,\bm k_1,\bm k_2, \bm k_3)$ is defined by
\begin{eqnarray}
\left\langle\delta(t,\bm k_1)\delta(t,\bm k_2)\delta(t,\bm k_3)\right\rangle
=(2\pi)^3\delta_D^{(3)}(\bm k_1+\bm k_2+\bm k_3)B_g(t,\bm k_1,\bm k_2, \bm k_3).
\end{eqnarray}
Thus the bispectrum $B_g(t,\bm k_1,\bm k_2, \bm k_3)$ relies on the implicit assumption ${\bm k}_1+{\bm k}_2+{\bm k}_3=0$.
This means that the bispectrum is described by the five parameters $k_1$, $k_2$,
$\cos\theta_{12}(={{\bm k}}_1\cdot{\bm k}_2/k_1k_2)$,
$\mu(=\cos\omega)$, and $\phi$ as variables,
with which we may write the vectors
\begin{eqnarray}
&&{\bm k}_1=(0,0,k_1),
\\
&&{\bm k}_2=(0,k_2\sin\theta_{12},k_2\cos\theta_{12}),
\\
&&{\bm k}_3=(0,-k_2\sin\theta_{12},-k_1-k_2\cos\theta_{12}),
\\
&&{\bm \gamma}=(\sin\omega\cos\phi,\sin\omega\sin\phi,\cos\omega),
\end{eqnarray}
where $\bm \gamma$ denotes the unit vector of the line of sight direction.
Figure~\ref{fig:configuration}  shows the configuration of the variables.
Then, we define $\mu_i$ as
\begin{eqnarray}
&&\mu_1={\hat{\bm k}}_1\cdot {\bm \gamma}=\cos\omega=\mu,
\\
&&\mu_2={\hat{\bm k}}_2\cdot {\bm \gamma}=
\sin\theta_{12}\sin\omega\sin\phi
+\cos\theta_{12}\cos\omega
=\sin\theta_{12}\sqrt{1-\mu^2}\sin\phi
+\cos\theta_{12}\mu,
\\
&&\mu_3=-{k_1\over k_3}\mu_1-{k_2\over k_3}\mu_2,
\end{eqnarray}
with ${k}_3^2=(k_2\sin\theta_{12})^2+(k_1+k_2\cos\theta_{12})^2$.
Hereafter, we use the notation $\theta=\theta_{12}$.
\def\thetaot{{\theta}}
Here we followed the choice of the variables introduced in Ref. \cite{Scoccimarro1999}.

The bispectrum in the halo approach consists of the one-halo term $B_{g,1h}$, the two-halo term $B_{g,2h}$,
and the three-halo term $B_{g,3h}$ given as
\begin{eqnarray}
B_g(t,\bm k_1,\bm k_2, \bm k_3)=B_{g,1h}(t,{\bm k}_1,{\bm k}_2,{\bm k}_3)
+B_{g,2h}(t,{\bm k}_1,{\bm k}_2,{\bm k}_3)
+B_{g,3h}(t,{\bm k}_1,{\bm k}_2,{\bm k}_3),
\end{eqnarray}
which are written as
\begin{eqnarray}
B_{g,1h}(t,{\bm k}_1,{\bm k}_2,{\bm k}_3)&=&{1\over \bar n^3}\int dM {dn(M)\over dM} \biggl[
\left<N_c\right>\left<N_s(N_s-1)\right>\left(\widetilde u({\bm k}_1,M)\widetilde u({\bm k}_2,M)
+2~{\rm cyclic~terms}\right)
\nonumber
\\
&&
+\left<N_s(N_s-1)(N_s-2)\right>\widetilde u({\bm k}_1,M)\widetilde u({\bm k}_2,M)\widetilde u({\bm k}_3,M)
\biggr],
\\
B_{g,2h}(t,{\bm k}_1,{\bm k}_2,{\bm k}_3)&=&{1\over \bar n^3}\int dM_1 {dn(M_1)\over dM_1} \biggl[\left<N_c\right>\left<N_s\right>\left(\widetilde u({\bm k}_1,M_1)+\widetilde u({\bm k}_2,M_1)\right)
\nonumber
\\
&&\hspace{2cm}
+\left<N_s(N_s-1)\right>\widetilde u({\bm k}_1,M_1)\widetilde u({\bm k}_2,M_1)\biggr]
\nonumber
\\
&&
\times\int dM_2 {dn(M_2)\over dM_2} \left(
\left<N_c\right>+\left<N_c\right>\left<N_s\right>\widetilde u({\bm k}_3,M_2)\right)P_{2h}(t,{\bm k}_3,M_1,M_2)
\nonumber
\\
&&\hspace{2cm}
+2~{\rm cyclic~terms},
\\
B_{g,3h}(t,{\bm k}_1,{\bm k}_2,{\bm k}_3)
&=&{1\over \bar n^3}\int \prod_{i=1}^3\left[dM_i {dn(M_i)\over dM_i}
\left<N_c\right>\left(1+\left<N_s\right>\widetilde u({\bm k}_i,M_i)\right)\right]
\nonumber
\\
&&~~~~~~~~~~~~~~~~~~~~~~\times
P_{3h}(t,{\bm k}_1,{\bm k}_2,{\bm k}_3,M_1,M_2,M_3),
\end{eqnarray}
  where $N_{\rm c}$ and $N_{\rm s}$ are the numbers of central galaxy and satellite galaxy,
  respectively, $\langle \cdots \rangle$ denotes the averaged value per halo with
fixing halo's mass under the assumption of the Poisson distribution,
$\langle N_{\rm c}\rangle$ and $\langle N_{\rm s}\rangle$ are defined by
Eqs. (\ref{defNc}) and (\ref{eq:HOD}), and we use the relations
$\langle N_{\rm s}(N_{\rm s}-1)\rangle=
\langle N_{\rm c}\rangle\langle N_{\rm s}\rangle^2$,
$\langle N_{\rm s}(N_{\rm s}-1)(N_{\rm s}-2)\rangle=
\langle N_{\rm c}\rangle\langle N_{\rm s}\rangle^3$ (see also \cite{YNH}),
$\bar{n}$ is the mean number density of galaxies given by
\begin{eqnarray}
\bar{n}=\int dM {dn\over dM} N_{\rm HOD}(M),
\end{eqnarray}
and we define
\begin{eqnarray}
&&\hspace{-1cm}P_{2h}(t,{\bm k}_3,M_1,M_2)=(b(M_1)+\mu_3^2 f)(b(M_2)+\mu_3^2 f)P_{m}^{\rm NL}(t,k_3),
\label{twohalom}
\\
&&\hspace{-1cm}P_{3h}(t,{\bm k}_1,{\bm k}_2,{\bm k}_3,M_1,M_2,M_3)
=2P_{m}^{\rm NL}(t,k_1)P_{m}^{\rm NL}(t,k_2)Z_1^{}({\bm k}_1,M_1)Z_1^{}({\bm k}_2,M_2)
Z_2^{}({\bm k}_1,{\bm k}_2,M_3)
\nonumber
\\
&&~~~~~~~~~~~~~~~~~
+2~{\rm cyclic~terms}
\label{threehalom}
\end{eqnarray}
with
\begin{eqnarray}
&&\hspace{-1cm}Z_1^{}({\bm k}_1,M_1)=b(M_1)+f\mu_1^2,
\\
&&\hspace{-1cm}Z_1^{}({\bm k}_2,M_2)=b(M_2)+f\mu_2^2,
\\
&&\hspace{-1cm}Z_2^{}({\bm k}_1,{\bm k}_2,M_3)=b(M_3)F_2({\bm k}_1,{\bm k}_2)+{b_2(M_3)\over 2}
+f\mu_{12}^2G_2({\bm k}_1,{\bm k}_2)
\nonumber\\
&&~~~~~~~~~~~~~
+{1\over2}f\mu_{12}k_{12}\left\{{\mu_1\over k_1}\left(b(M_3)+f\mu_2^2\right)
+{\mu_2\over k_2}\left(b(M_3)+f\mu_1^2\right)\right\},
\end{eqnarray}
$\mu_{12}=({\bm k}_1+{\bm k}_2)\cdot {\bm \gamma}/k_{12}$, and $k_{12}=|{\bm k}_1+{\bm k}_2|$,
and where $P_m^{\rm NL}(t,k)$ is the matter power spectrum at time $t$, for which we use the
nonlinear fitting formula for the matter power spectrum \cite{NFWprofile}.
We also use the fitting formula of the linear growth rate
$f={d\log D_1(a)/d\log a}=[\Omega_m(a)]^\gamma$,
where $\Omega_m(a)$ is the matter density parameter at the scale
factor $a=a(t)$ and $\gamma=0.55$.
For the linear bias $b(M)$, we adopt the halo bias of the fitting function,
\begin{eqnarray}
b(M)=1-{\nu^a\over \nu^a+\delta_c^a}+0.183\nu^b+0.265 \nu^c,
\label{biasmodel}
\end{eqnarray}
with $a = 0.132$, $b = 1.5$, and $c = 2.4$, which was calibrated using $N$-body simulations
\cite{Tinker}.

\section{Multipole bispectrum}

\subsection{Definition of multipole bispectrum}

The bispectrum in redshift space is specified by 5 parameters, for which
we adopt $k_1,~k_2,~\theta_{12},~\omega,~\phi$, as is described in the previous section II.B,
following Ref.~\cite{Scoccimarro1999}.
The parameters $\omega$ and $\phi$ take the values $0\leq\omega\leq\pi$ and
$0\leq\phi\leq 2\pi$. Then, we consider the multipole expansion of the
bispectrum in terms of spherical harmonics, which is usually defined
as
\begin{eqnarray}
  \label{ylm}
  Y_\ell^m(\omega,\phi)=i^{m+|m|}\sqrt{\frac{(2\ell+1)}{4\pi}\frac{(\ell-|m|)!}
  {(\ell+|m|)!}}P_\ell^{|m|}(\cos{\omega})e^{im\phi},
\end{eqnarray}
where $m$ is an integer in the range $-\ell \leq m \leq \ell$, and the associated
 Legendre polynomials are defined by
\begin{eqnarray}
  \label{legendre}
  &&P_\ell^{|m|}(\mu)=(1-\mu^2)^{|m|/2}\frac{d^{|m|}}{d\mu^{|m|}}P_\ell(\mu).
\end{eqnarray}
Note that $P_\ell^{|m=0|}(\mu)$ reduces to the Legendre polynomial $P_\ell(\mu)$.

We adopt the spherical harmonics as a set of real functions, and we define
\begin{eqnarray}
&&Y_{\ell,m,c}(\omega,\phi)=
\left\{
\begin{array}{ll}
      \sqrt{{2\ell+1\over 4\pi}}P_\ell(\cos\omega)&~~~~~(m=0),
\\
      (-1)^{(m+|m|)/2}\sqrt{\frac{(2\ell+1)}{4\pi}
      \frac{(\ell-|m|)!}{(\ell+|m|)!}}P_\ell^{|m|}(\cos\omega)\sqrt{2}\cos{m\phi}& ~~~~~(m\neq0),
\end{array}
\right.
  \label{realYlmsig}
\\
&&Y_{\ell,m,s}(\omega,\phi)=(-1)^{(m+|m|)/2}\sqrt{\frac{(2\ell+1)}{4\pi}
      \frac{(\ell-|m|)!}{(\ell+|m|)!}}P_\ell^{|m|}(\cos\omega)\sqrt{2}\sin{m\phi} ~~~~(m\neq0).
\label{realYlmsig2}
\end{eqnarray}
These functions satisfy the normalization
\begin{eqnarray}
\int_0^{2\pi}d\phi\int_{0}^{\pi} d\omega\sin\omega Y_{\ell,m,\sigma}(\omega,\phi)Y_{\ell',m',\sigma'}(\omega,\phi)
=\delta_{\ell\ell'}\delta_{mm'}\delta_{\sigma\sigma'},
\label{YYnorm}
\end{eqnarray}
where $\sigma$ and $\sigma'$ denote $c$ or $s$, though we excluded
$Y_{\ell,0,s}(\omega,\phi)$ because it is zero, which is not defined.

Now we define the multipoles of the bispectrum by
\begin{eqnarray}
    \label{Blmsig}
    B^{\ell,m,\sigma}(k_1,k_2,\theta) = \sqrt{1\over {4\pi}(2\ell+1)}\int_0^{2\pi}
    d\phi \int_{-1}^{+1}d\cos \omega B_g(t,k_1,k_2,\theta,\omega,\phi){Y}_{\ell,m,\sigma}(\omega,\phi).
\end{eqnarray}
The reduced bispectrum is defined in a similar way to what was done the previous work \cite{YNH}:
\begin{eqnarray}
    \label{Qlmsig}
    Q^{\ell,m,\sigma}(k_1,k_2,\theta)
    ={B^{\ell,m,\sigma}(k_1,k_2,\theta) \over P^0(t,k_1)P^0(t,k_2)+P^0(t,k_2)P^0(t,k_3)+P^0(t,k_3)P^0(t,k_1)},
\end{eqnarray}
where $P^0(t,k_i)$ is the monopole spectrum of the galaxy power spectrum $P_g(t,\bm k_i)$, i.e.,
\begin{eqnarray}
P^0(t,k_i)={1\over 2}\int_{-1}^{+1}d\mu P_{g}(t,{\bm k}_i).
\end{eqnarray}
In our modeling on the basis of the halo approach,  $P_g(t,\bm k_i)$ is obtained by
a combination of the one-halo term and the two-halo term \cite{HY}:
\begin{eqnarray}
P_g(t,\bm k)=P_{g,1h}(t,{\bm k})+P_{g,2h}(t,{\bm k}),
\label{Pgstk}
\end{eqnarray}
where we defined
\begin{eqnarray}
&&P_{g,1h}(t,{\bm k})={1\over {\bar n}^2}\int dM {dn\over dM}
\left[2\left<N_c\right>\left<N_s\right>\widetilde u({\bm k},M)
+\left<N_s(N_s-1)\right>\widetilde u^2({\bm k},M)\right],
\label{onehaloterm}
\\
&&P_{g,2h}(t,{\bm k})={1\over {\bar n}^2}\prod_{i=1}^2\left[\int dM_i {dn\over dM_i}
\left<N_c\right>\left\{1+\left<N_s\right>\widetilde u({\bm k},M_i)\right\}(b(M_i)+f\mu^2)
\right]P_m^{\rm NL}(t,k).
\nonumber\\
\label{twohaloterm}
\end{eqnarray}
Definitions (\ref{Blmsig}) and (\ref{Qlmsig}) reduce to those in the previous paper \cite{YNH} when $m=0$.
Following the definition of the spherical harmonics (\ref{realYlmsig}) and (\ref{realYlmsig2}),
we list the explicit expression for the case $\ell \leq 4$ in Appendix A.

Since the bispectrum consists of a one-halo term, a two-halo term, and a three-halo term, we can express the total bispectrum
as the sum of corresponding halo terms:
\begin{eqnarray}
  B^{\ell,m,\sigma}(k_1,k_2,\theta)=  B^{\ell,m,\sigma}_{1h}(k_1,k_2,\theta) + B^{\ell,m,\sigma}_{2h}(k_1,k_2,\theta) + B^{\ell,m,\sigma}_{3h}(k_1,k_2,\theta),
\end{eqnarray}
and the reduced total bispectrum is
\begin{eqnarray}
  \label{totalQ}
  Q^{\ell,m,\sigma}(k_1,k_2,\theta)=  Q^{\ell,m,\sigma}_{1h}(k_1,k_2,\theta) + Q^{\ell,m,\sigma}_{2h}(k_1,k_2,\theta)
  + Q^{\ell,m,\sigma}_{3h}(k_1,k_2,\theta).
\end{eqnarray}

\begin{table}[t]
\begin{center}
\begin{tabular}{|lccc|}
\hline
\vspace{0mm}
~~~{$B^{\ell,m(,\sigma)}$}~ & ~$B_{1h}^{\ell,m,\sigma}$~& ~$B_{2h}^{\ell,m,\sigma}$~& ~$B_{3h}^{\ell,m,\sigma}$~\\
\hline
~~~$B^{0,0,c}$  & $\bullet$ & $\bullet$ & $\bullet$
\vspace{-0mm}
\\
~~~$B^{2,0,c}$  & $\bullet$ & $\bullet$ & $\bullet$
\vspace{-0mm}
\\
~~~$B^{2,1,s}$  & $\circ$ & $\circ$ &  $\circ$
\vspace{-0mm}
\\
~~~$B^{2,2,c}$  & $\circ$ & $\circ$ &  $\circ$
\vspace{-0mm}
\\
~~~$B^{4,0,c}$  & $\bullet$ & $\bullet$ & $\bullet$
\vspace{-0mm}
\\
~~~$B^{4,1,s}$  & $\circ$ & $\circ$ &  $\circ$
\vspace{-0mm}
\\
~~~$B^{4,2,c}$  & $\circ$ &  $\circ$ & $\circ$
\vspace{-0mm}
\\
~~~$B^{4,3,s}$  & $\circ$ & $\circ$ &  $\circ$
\vspace{-0mm}
\\
~~~$B^{4,4,c}$  & $\circ$ & $\circ$ & $\circ$
\\
\hline
\end{tabular}
\caption{Nonvanishing multipoles of the bispectrum $B^{\ell,m,\sigma}$ up to $\ell=4$.
$\bullet$ means a nonvanishing quantity in previous work \cite{YNH},
while $\circ$ means one found in this work.}
\label{table:nozeromultipoles}
\end{center}
\end{table}

\subsection{Results}

We demonstrate the characteristic behaviors of the multipole bispectrum defined in the
previous subsection.
There are nine nonvanishing components of $B^{\ell,m,\sigma}$ up to $\ell=4$
(see Table \ref{table:nozeromultipoles}), of which $3$ multipoles with
$m=0$ have been known so far \cite{Scoccimarro1999,Scoccimarro2015};
these we investigated in our previous work \cite{YNH}, while six multipoles
denoted by the symbol $\circ$ in the table are the new components, which we investigated in
the present paper.
The other components up to $\ell=4$ are zero, because of the symmetry with respect to $\phi$.

Figure 2 shows the characteristic behaviors of the new nonzero components,
where we adopted the HOD parameters of the LOWZ sample in Table I.
Each panel of Fig. 2 plots $Q^{2,1,s}$, $Q^{2,2,c}$, $Q^{4,1,s}$, $Q^{4,2,c}$, $Q^{4,3,s}$, and $Q^{4,4,c}$ as functions of
$\theta$ with $k_1$ and $k_2=2k_1$ fixed. Each multipole bispectrum shows unique behaviors.
One can see that the one-halo term (green dotted curve)
and the two-halo term (blue dashed curve) make significant contributions to these multipole bispectrum
and dominate over the contribution from the three-halo term (red long-dashed curve) for the case
$k_1>0.1{\rm ~Mpc}^{-1}$. This is significant for
$Q^{4,1,s}$, $Q^{4,2,c}$, $Q^{4,3,s}$, and $Q^{4,4,c}$ than $Q^{2,1,s}$ and $ Q^{2,2,c}$.

The contributions of the two-halo term and the one-halo term are opposite compared with the three-halo
term for $Q^{2,1,s}$ and $Q^{2,2,c}$. This is also true for $Q^{2,0,c}$ investigated in the previous work \cite{YNH}.
This can be understood as follows: The higher multipole bispectrum reflects the
redshift space distortions. The contributions of the two-halo term and the one-halo term
reflect the FoG effect, while the three-halo term contribution reflect the linear distortion.
These two redshift-space distortions have an opposite effect in the quadrupole power spectrum and bispectrum.

  \begin{figure}[H]
   \begin{minipage}{0.45\hsize}
    \begin{center}
    \includegraphics[width=6.5cm,height=6.5cm]{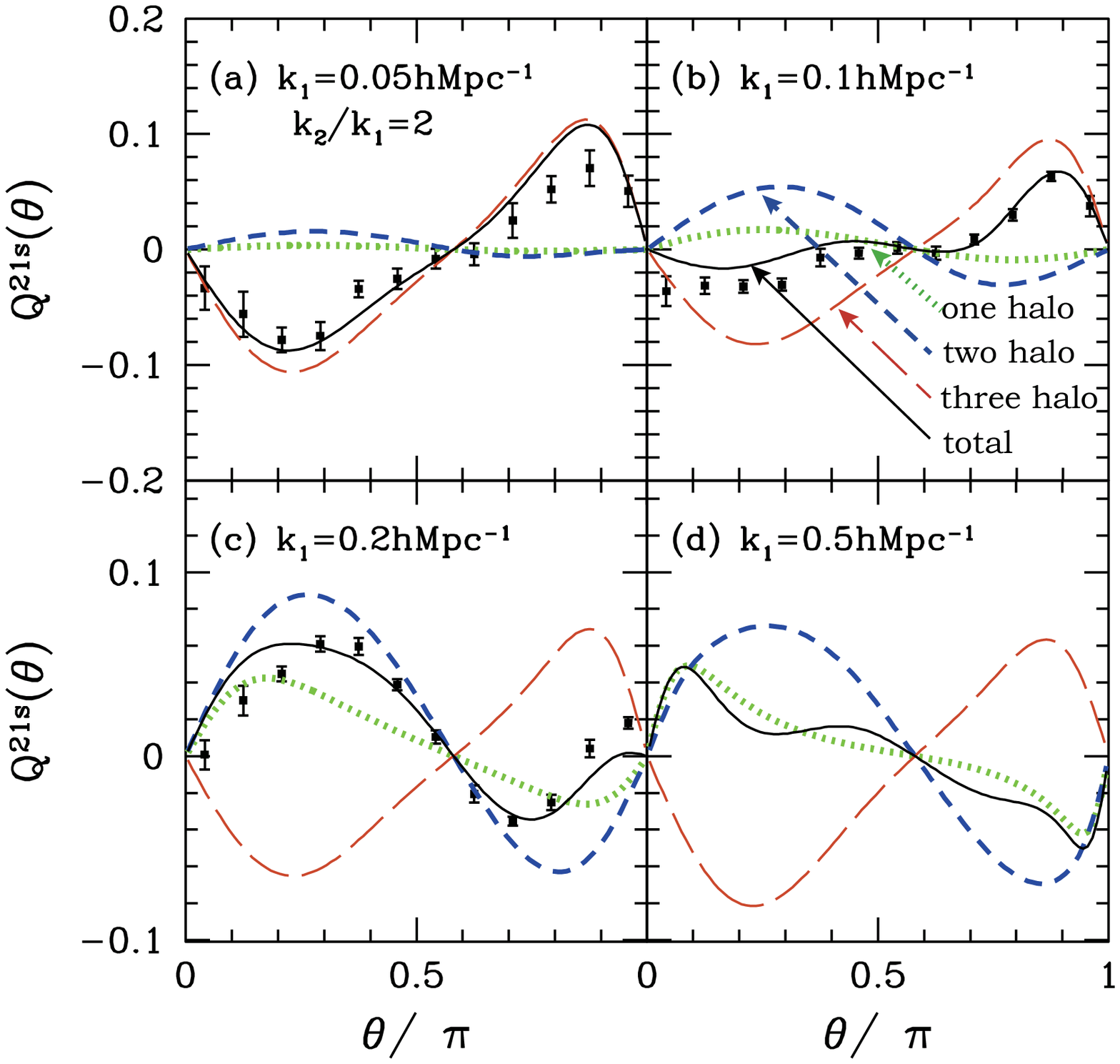}
    \end{center}
     \vspace{-0.cm}
    \label{fig:one}
   \end{minipage}\begin{minipage}{0.45\hsize}
    \begin{center}
    \includegraphics[width=6.5cm,height=6.5cm]{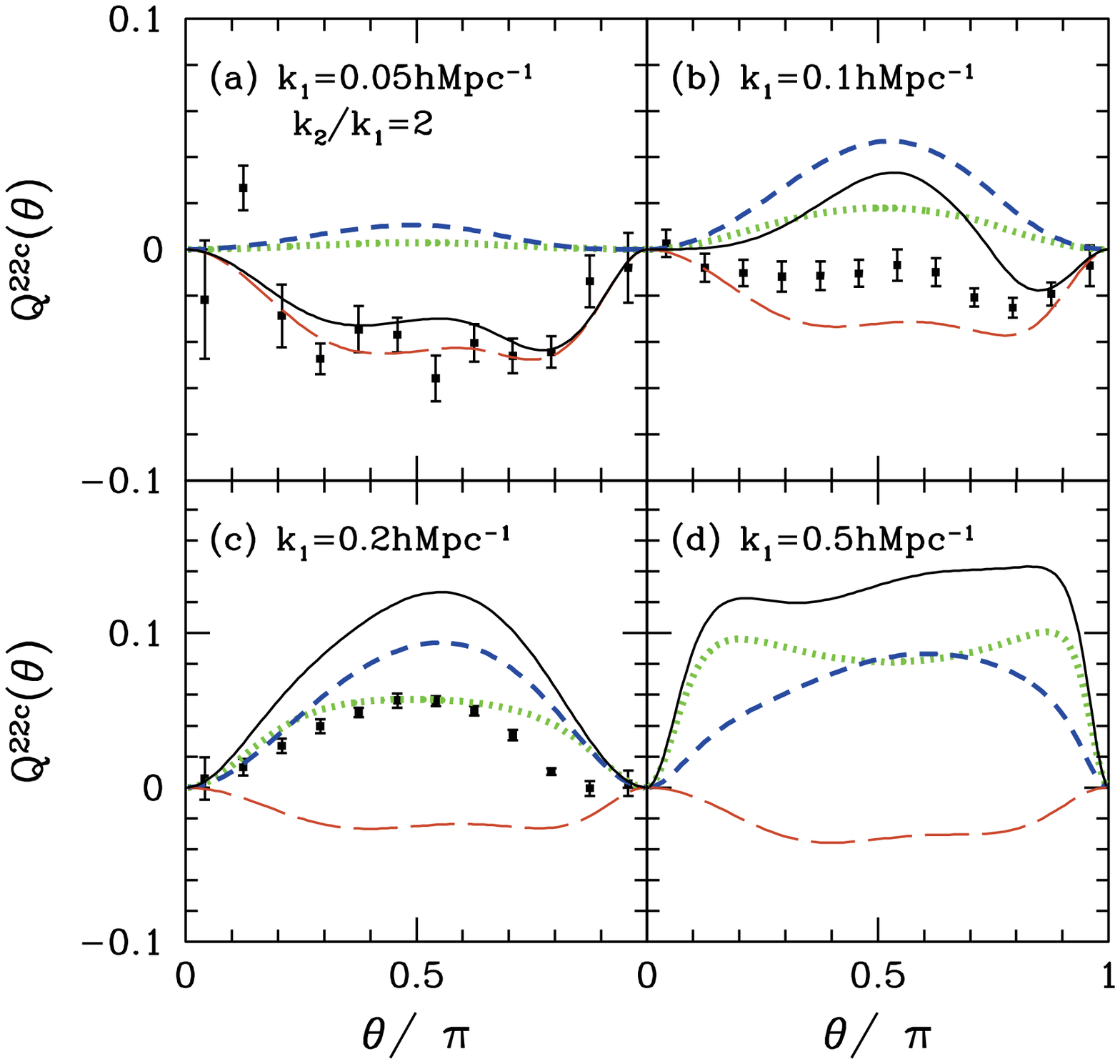}
    \end{center}
     \vspace{-0.cm}
    \label{fig:two}
   \end{minipage}
  \begin{minipage}{0.45\hsize}
    \begin{center}
    \includegraphics[width=6.5cm,height=6.5cm]{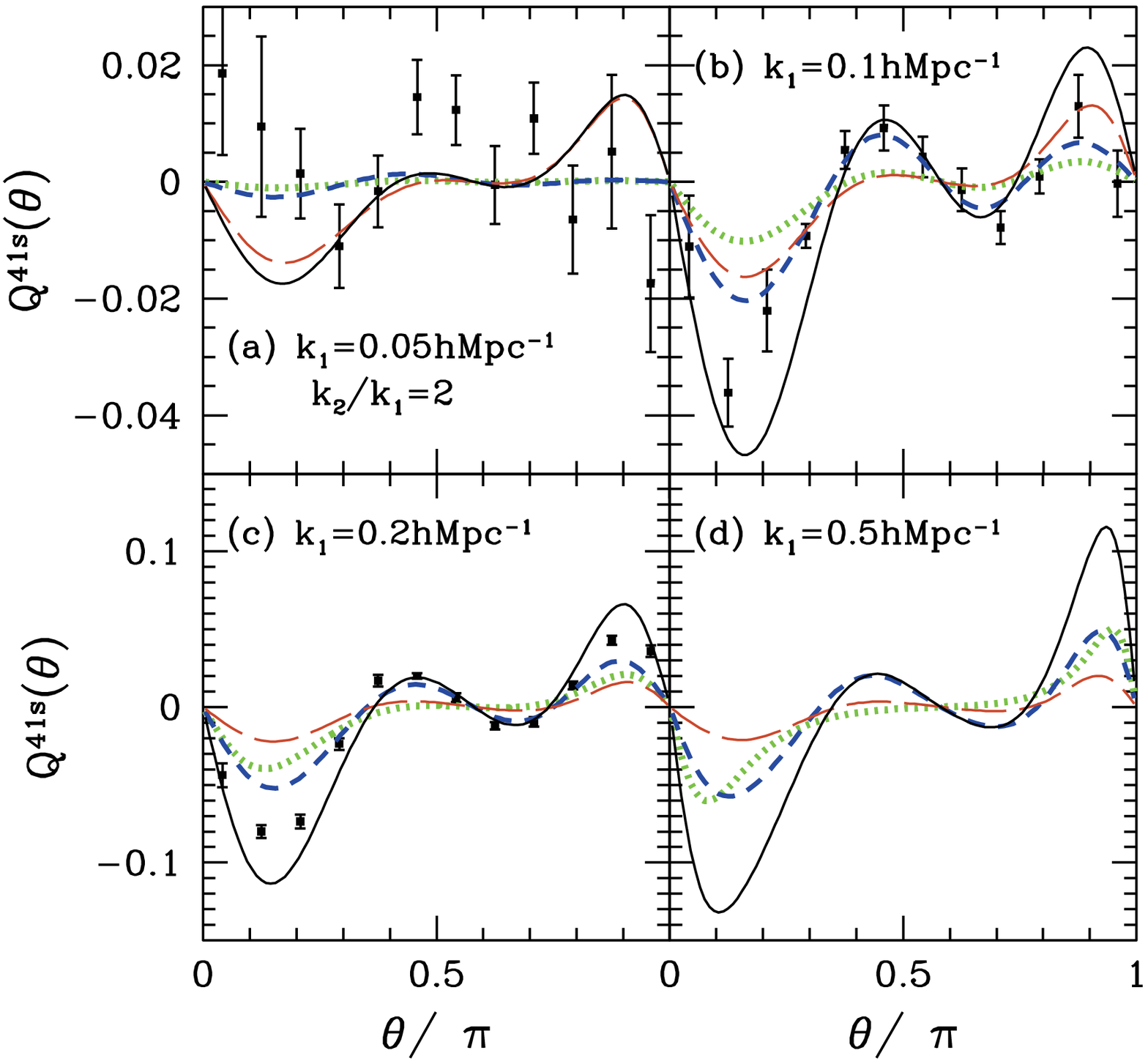}
    \end{center}
     \vspace{-0.cm}
    \label{fig:one}
   \end{minipage}\begin{minipage}{0.5\hsize}
    \begin{center}
    \includegraphics[width=6.5cm,height=6.5cm]{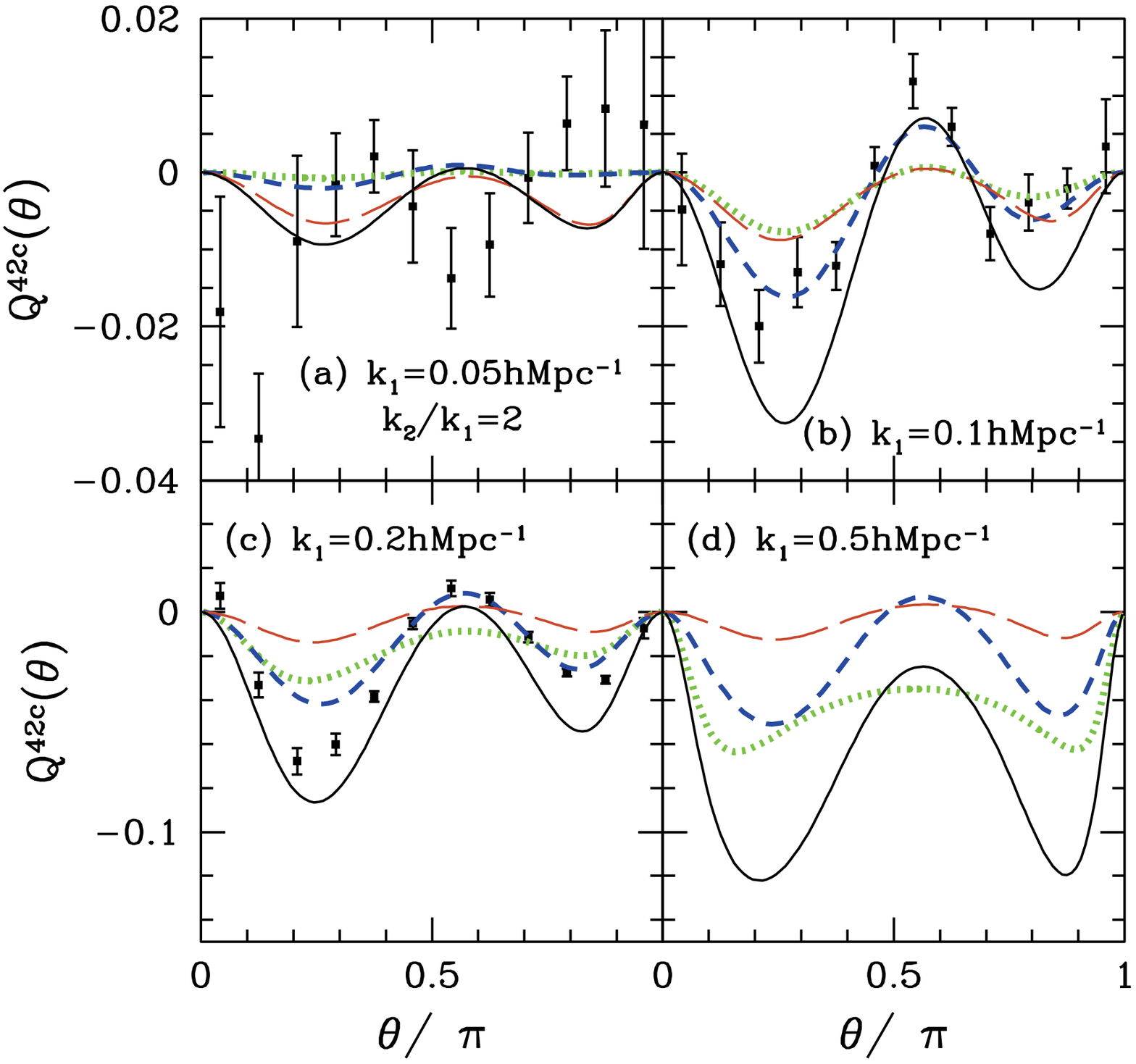}
    \end{center}
     \vspace{-0.cm}
    \label{fig:two}
   \end{minipage}
   \begin{minipage}{0.45\hsize}
    \begin{center}
    \includegraphics[width=6.5cm,height=6.5cm]{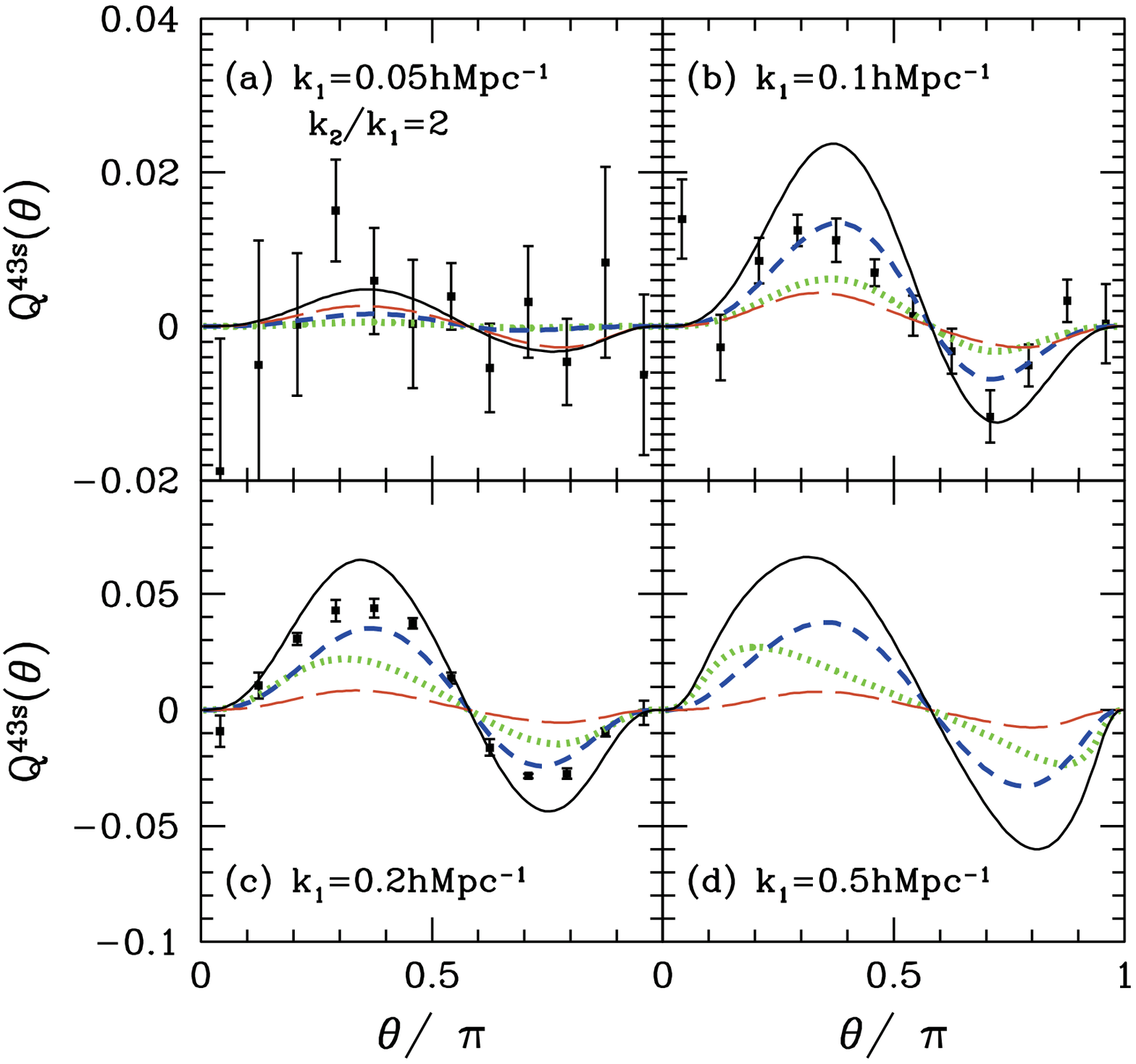}
    \end{center}
     \vspace{-0.cm}
    \label{fig:three}
   \end{minipage}\begin{minipage}{0.5\hsize}
    \begin{center}
    \includegraphics[width=6.5cm,height=6.5cm]{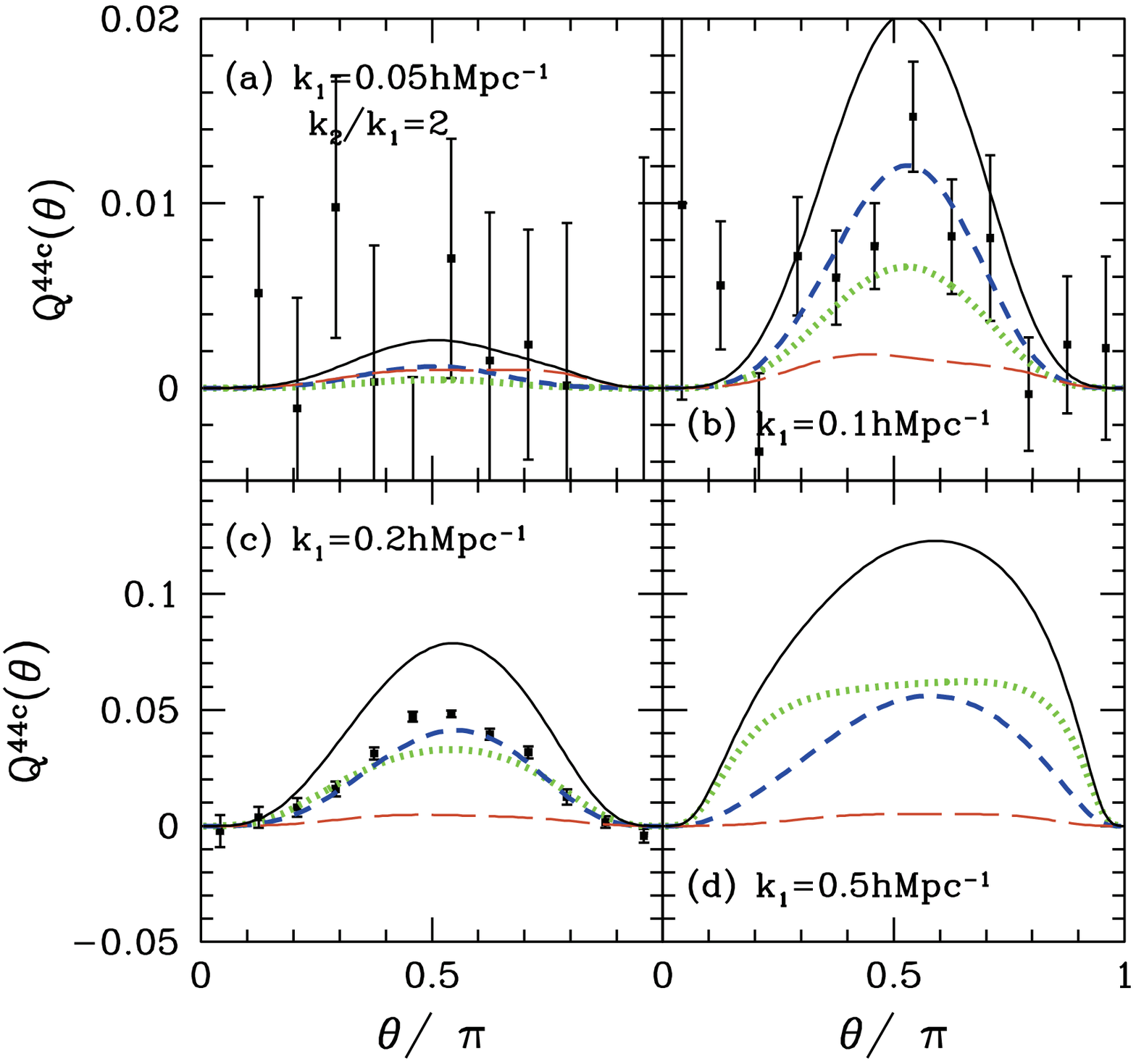}
    \end{center}
     \vspace{-0.cm}
  \label{fig:four}
   \end{minipage}
  \caption{Nonvanishing multipoles of the reduced bispectrum, $Q^{2,1,s}$, $Q^{2,2,c}$, $Q^{4,1,s}$, $Q^{4,2,c}$, $Q^{4,3,s}$, and $Q^{4,4,c}$
  as function of $\theta$ for the LOWZ sample by fixing (a) $k_1=0.05$,
  (b) $k_1=0.1$, (c) $k_1=0.2$, and (d) $k_1=0.5$ in units of $h$/Mpc
  and $k_2/k_1=2$. In each panel the (green) dotted curve is the one-halo term contribution,
  the (blue) short-dashed curve is the two-halo term contribution, the (red) long-dashed curve is
  the three-halo term contribution, and the (black) solid curve is the total combination.
  Here we set $b_2=0$.
  The data points with error bars show the results of the mock catalogs.
  Result of the mock catalog for the case $k_1=0.5$ is not available due to
    lack of the resolution of our simulation.
  \label{figurenonzero}
  }
  \end{figure}

\subsection{Comparison with the results of mock catalogs}
We compare our analytic model with the results of mock catalogs by
assuming the HOD of the SDSS-III BOSS LOWZ sample.
A similar comparison was done for $Q^{0,0}$, $Q^{2,0}$, and $Q^{4,0}$ in our previous paper \cite{YNH},
which is also adopted for comparison in the present paper.
We run 10 realizations of $N$-body simulations at a side length of $1h^{-1}$ Gpc
with the number of mass particles set as 800$^3$ (where the mass for each
particle is set as $1.3\times 10^{11}h^{-1}\;M_\odot$) using the Gadget-2 code
\cite{Springel05}. The softening length is set to be $50h^{-1}$ kpc.
The initial mass distribution is Gaussian, starting
from $z=49$ generated by the 2LPT code of \cite{Crocce06}.
The halo is identified with
the friends-of-friends algorithm with a linking length of 0.2. The
minimum number of mass particles is 10, corresponding to a mass of
$1.3\times 10^{12}h^{-1}\;M_\odot$. The
central and satellite galaxies are assigned to each halo to follow the
HOD of the BOSS LOWZ sample. The position and velocity of each central
galaxy are given as the arithmetic mean of all particles in the
halo. The position and velocity of satellites are defined as those of
randomly selected mass particles.  We confirmed that the mass
resolution of our simulation is sufficient for the following
comparison with our theoretical model.

The data points with error bars in Fig.~\ref{figurenonzero} show the result of the mock catalogs.
The error bars represent the one-sigma dispersion of
$10$ simulation results divided by $\sqrt{10}$, which roughly
corresponds to the sample variance for $10\;({\rm Gpc}/h)^3$ volume
data.
As is demonstrated in \cite{YNH} for $Q^{0,0,c}$, $Q^{2,0,c}$, and $Q^{4,0,c}$, our theoretical model well explains
the characteristic behavior of the bispectrum from the mock catalogs even for
$Q^{\ell,m,\sigma}$ with $m\neq0$, though some
differences arise for the cases with larger wavenumbers at a quantitative level.
However, the behaviors of the simulations are reproduced at a qualitative level.
  Behaviors of the galaxy bispectrum at large wavenumbers have not been well studied.
  One of the reason might be the galaxy bispectrum sensitively depends on the HOD
  parameter as our result suggests.
  Even for the halo bispectrum, we don't know a precise formula which reproduces
  the three halo term being valid at higher wavenumbers.
  Our analytic model of the three halo term is the simplest model based on the
  lowest order of density perturbations, which must be improved for comparison
  with mock catalogs or observations in future.

\section{Discussion: approximate formulas}
In this section, we consider approximate formulas, which roughly explain the characteristic
behaviors of the multipoles of the bispectrum.
Since these formulas are too long to be fully presented in the main part, they are listed in Appendix B.
These approximate formulas are useful for understanding the behaviors of the multipole bispectrum.

According to previous works \cite{HY,YNH}, we may introduce the following approximate formulas
for the one-halo term, the two-halo term, and the three-halo term:
\begin{eqnarray}
B_{g,1h}(t,{\bm k}_1,{\bm k}_2,{\bm k}_3)&\simeq&{f_s^2\over \bar n^2}
\left(\widetilde u({\bm k}_1)\widetilde u({\bm k}_2)+\widetilde u({\bm k}_2)\widetilde u({\bm k}_3)+\widetilde u({\bm k}_3)\widetilde u({\bm k}_1)\right),
\\
B_{g,2h}(t,{\bm k}_1,{\bm k}_2,{\bm k}_3)
&\simeq&{f_s\over \bar n}
\left(\widetilde u({\bm k}_1)+\widetilde u({\bm k}_2)\right)(\bar b+\mu_3^2f)^2P_m^{\rm NL}(k_3)
+2~{\rm cyclic~terms},
\\
B_{g,3h}(t,{\bm k}_1,{\bm k}_2,{\bm k}_3)&\simeq&
2P_m^{\rm NL}(t,k_1)P_m^{\rm NL}(t,k_2)(\bar b+\mu_1^2f)(\bar b+\mu_2^2f)
\Bigl[\bar bF_2({\bm k}_1,{\bm k}_2)+f \mu_{3}^2G_2({\bm k}_1,{\bm k}_2) +{\bar b_2\over 2}
\nonumber\\
&&-{1\over 2}f\mu_3k_3
\Bigl({\mu_1\over k_1}(\bar b+f\mu_2^2)+{\mu_2\over k_2}(\bar b+f\mu_1^2)\Bigr)\Bigr]
+2~{\rm cyclic~terms},
\end{eqnarray}
respectively, where we use the approximate formula
\begin{eqnarray}
&&\widetilde u({\bm k_i})\simeq 
\exp\left[-\frac{\overline\sigma_{v,{\rm off}}^2k_i^2\mu^2}
{2a^2H^2(z)}\right]=\exp\left[-{\lambda^2k_i^2\mu^2}\right],
\end{eqnarray}
for $i=1,~2$, and $3$, where $\bar b$ and $\overline \sigma_{v,{\rm off}}$ are
averaged values of the bias and the random velocity of satellite galaxies
over the halo mass and $f_s$ is the satellite fraction.
Here we introduce the characteristic length scale, associated with the
random motions by
\begin{eqnarray}
  \lambda^2
  =\frac{\overline\sigma_{v,{\rm off}}^2}
{2a^2H^2(z)},
\end{eqnarray}
     where $aH(z)$ is the value at the mean redshift $\bar z$ as
     $aH(z)=H(\bar z)/(1+\bar z)$.

Then, we may write the approximate formula for the multipole bispectrum in the form expressing
the dependence on $f_s/\bar{n}$ explicitly as
\begin{equation}
  B^{\ell,m,\sigma}(k_1,k_2,\theta)=  \frac{f_s^2}{\bar{n}^2} \widetilde{B}^{\ell,m,\sigma}_{1h}(k_1,k_2,\theta)
  + \frac{f_s}{\bar{n}} \widetilde{B}^{\ell,m,\sigma}_{2h}(k_1,k_2,\theta) + \widetilde{B}^{\ell,m,\sigma}_{3h}(k_1,k_2,\theta),
\end{equation}
where the formulas of $\widetilde{B}^{\ell,m,\sigma}$ are presented in Appendix B.

The mathematical formulas of Appendix B are derived using {\it Mathematica}. The
source {\it Mathematica } programs for the derivation are provided in the source
file of arXiv:1706.03515.

From the approximation formulas, in general, we have
\begin{eqnarray}
  &&B_{1h}^{\ell,m,\sigma} \sim \left(\frac{f_s}{\bar{n}}\right)^2 (\lambda k)^\ell \sin^m \theta_{12},
  \label{1hfactor}
\\
   &&B_{2h}^{\ell,m,\sigma} \sim  \frac{f_s}{\bar{n}}  \left[\left({\cal O}(b^2 P_m)
    +{\cal O}(b f P_m)+{\cal O}(f^2 P_m)\right) + {\cal O} (\lambda^2 k^2 P_g ) \right] \sin^m \theta_{12},
    \label{2hfactor}
\\
   &&B_{3h}^{\ell,m,\sigma} \sim f^{\ell/2} {\cal O}(P_m^2) \sin^m \theta_{12}.
   \label{3hfactor}
\end{eqnarray}
The factor $\sin^m \theta_{12}$ comes from the mathematical properties of spherical harmonics.

For the case $\ell=0$, as discussed in Ref.~\cite{YNH}, $B^{0,0,c}_{1h}(k_1,k_2,\theta)\simeq\frac{f_s^2}{\bar{n}^2}
(3-{\cal O}(k^2\lambda^2))$. Thus, Eq.~(\ref{1hfactor}) means that the one-halo term
makes a contribution to the multipole bispectrum dominantly from the term in proportion to
$(k\lambda)^\ell$, which comes from the FoG effect.
Equation~(\ref{3hfactor}) reveals that the three-halo term is in proportion to
the factor $f^{\ell/2}$, where $f$ denotes the linear growth rate, which shows that the
three-halo term contribution comes from the linear redshift-space distortion.
The contribution from the two-halo term includes both  the FoG effect and the
linear redshift-space distortion effect depending on the scales $k$,  from Eq.~(\ref{2hfactor}).
In the two-halo term, the FoG effect is the dominant contribution for scales larger than $k\sim 0.1\;{\rm Mpc}^{-1}$.
The total combination of the one-halo term, the two-halo term, and the three-halo term contributes to the complicated behaviors of the multipole bispectrum.

\section{Summary and Conclusions}

In this work, as a generalization of the halo approach to the galaxy bispectrum in redshift space,
we found six new nonvanishing multipole components up to $\ell=4$.
We demonstrated the characteristic behaviors of these nonvanishing multipoles, assuming the HOD
parameters of the LOWZ sample. Each component shows unique behaviors.
Using an analytic approximate method, we investigated how the one-halo term,
the two-halo term, and the three-halo term make contributions to the multipole bispectrum.
This has revealed that the higher multipole bispectrum is significantly contaminated by the
FoG effect on scales larger than $k\sim 0.1~{\rm Mpc}^{-1}$.
The total bispectrum is determined by the balance between the contributions of the FoG effect
and the linear redshift-space distortion and is complicated.
This study shows that the bispectrum reflects the cosmological model
and the physical properties of the galaxy sample. These properties are interesting because
we might be able to test the cosmological model as well as the galaxy--halo connection,
leading to better understandings of the large-scale structure formation.

A more precise theoretical model will be necessary to extract cosmological information
from observational data of ongoing and future galaxy redshift surveys (e.g., SDSS IV,
Subaru/PFS, and EUCLID), which will be a necessary investigation as an extension to our model.
Because the validity of standard perturbation theory at leading order is somehow limited to the linear regime,
this limits the validity of $P_{2h}(t,k_1,k_2,M_1,M_2)$ and $P_{3h}(t,k_1,k_2,k_3,M_1,M_2,M_3)$.
More precise evaluation of the nonlinearity
including the next-to-leading order correction and higher order corrections is needed
for further extension. As an example, in Ref. \cite{HST}, a model for the matter
bispectrum considering one-loop order correction in redshift space has been developed
by introducing a univariate function with a single free parameter.
Improvement of modeling the correlation of halos by including the higher order
corrections will be necessary in the future.

\section*{Acknowledgment}
This work was supported by MEXT/JSPS KAKENHI Grant Numbers 15H05895 and JP16H03977.
We thank A. Taruya, I. Hashimoto, Y. Rasera, T. Nishimichi, N. Yoshida, M. Takada,
and N. Sugiyama for useful comments.

\appendix
\section{Spherical Harmonics}
{\small
\begin{eqnarray*}
&&\hspace{-1cm}Y_{0,0}={1\over \sqrt{4\pi}},~~
Y_{1,0}=\sqrt{3\over {4\pi}}\cos\omega,~~
Y_{1,1,c}=-\sqrt{3\over {4\pi}}\sin\omega\cos\phi, ~~Y_{11s}=-\sqrt{3\over {4\pi}}\sin\omega\sin\phi, ~~
\\
&&\hspace{-1cm}Y_{2,0}=\sqrt{5\over {16\pi}}(3\cos^2\omega-1),~~
Y_{2,1,c}=-\sqrt{15\over {4\pi}}\sin\omega\cos\omega\cos\phi, ~~
  Y_{2,1,s}=-\sqrt{15\over {4\pi}}\sin\omega\cos\omega\sin\phi,
\\
&&\hspace{-1cm}Y_{2,2,c}=+\sqrt{15\over {16\pi}}\sin^2\omega \cos 2\phi, ~~
  Y_{2,2,s}=+\sqrt{15\over {16\pi}}\sin^2\omega \sin 2\phi, ~~
Y_{3,0}=\sqrt{7\over {16\pi}}(5\cos^3\omega-3\cos\omega),~~
\\
&&\hspace{-1cm}Y_{3,1,c}=-\sqrt{21\over {32\pi}}\sin\omega(5\cos^2\omega-1)\cos\phi, ~~
  Y_{3,1,s}=-\sqrt{21\over {32\pi}}\sin\omega(5\cos^2\omega-1)\sin\phi, ~~
\\
&&\hspace{-1cm}Y_{3,2,c}=+\sqrt{105\over {16\pi}}\sin^2\omega\cos\omega\cos 2\phi, ~~
  Y_{3,2,s}=+\sqrt{105\over {16\pi}}\sin^2\omega\cos\omega\sin 2\phi, ~~
\\
&&\hspace{-1cm}Y_{3,3,c}=-\sqrt{35\over {32\pi}}\sin^3\omega\cos 3\phi, ~~
  Y_{3,3,s}=-\sqrt{35\over {32\pi}}\sin^3\omega\sin 3\phi, ~~
\\
&&\hspace{-1cm}
Y_{4,0}=\sqrt{9\over {256\pi}}(35\cos^4\omega-30\cos^2\omega+3),~~
\\
&&\hspace{-1cm}Y_{4,1,c}=-\sqrt{45\over {32\pi}}\sin\omega(7\cos^2\omega-3)\cos\omega\cos\phi, ~~
  Y_{4,1,s}=-\sqrt{45\over {32\pi}}\sin\omega(7\cos^2\omega-3)\cos\omega\sin\phi, ~~
\\
&&\hspace{-1cm}Y_{4,2,c}=+\sqrt{45\over {64\pi}}(-7\cos^4\omega+8\cos^2\omega-1)\cos 2\phi, ~~
  Y_{4,2,s}=+\sqrt{45\over {64\pi}}(-7\cos^4\omega+8\cos^2\omega-1)\sin 2\phi, ~~
\\
&&\hspace{-1cm}Y_{4,3,c}=-\sqrt{315\over {32\pi}}\sin^3\omega\cos\omega\cos 3\phi, ~~
  Y_{4,3,s}=-\sqrt{315\over {32\pi}}\sin^3\omega\cos\omega\sin 3\phi, ~~
\\
&&\hspace{-1cm}Y_{4,4,c}=+\sqrt{315\over {256\pi}}\sin^4\omega\cos 4\phi, ~~
  Y_{4,4,s}=+\sqrt{315\over {256\pi}}\sin^4\omega\sin 4\phi, ~~
\end{eqnarray*}

\section{Approximate formulas}
  \begin{eqnarray}
    &&\hspace{-1cm}\widetilde{B}_{1h}^{2,1,s} = \frac{1}{1155 \sqrt{3}}\lambda ^2 k_2 \sin \theta_{12}
       (k_1+2 k_2 \cos \theta_{12})\bigg\{924-594 \lambda ^2 \Big(k_1^2+k_1 k_2 \cos \theta_{12}+k_2^2\Big)\notag\\[5pt]
    &&\hspace{-1cm}~~
       +33\lambda ^4 \Big[5 k_1^4+k_1 k_2 \big(5 \cos \theta_{12} (k_1^2+2k_2^2)+3 k_1 k_2 \cos 2\theta_{12}\big)+7 k_1^2 k_2^2+5 k_2^4\Big]\notag\\[5pt]
    &&\hspace{-1cm}~~
       -5 \lambda ^6 \Big[7 k_1^6+5k_1^4 k_2^2+12 k_1^2 k_2^4+k_1 k_2 \Big(k_1 k_2\big(\cos 2\theta_{12} (2 k_1^2+9
       k_2^2)+k_1 k_2 \cos 3 \theta_{12}\big)+\notag\\[5pt]
    &&\hspace{-1cm}~~
       \qquad \cos \theta_{12}(7 k_1^4+6 k_1^2 k_2^2+21 k_2^4)\Big)+7k_2^6)\Big]\bigg\}
  \end{eqnarray}
  \begin{eqnarray}
       &&\hspace{-1cm}\widetilde{B}_{1h}^{2,2,c} = \frac{1}{2310 \sqrt{3}}\lambda ^2 k_2^2 \sin ^2\theta_{12} \bigg\{1848
         -594 \lambda ^2 \Big(k_1^2+2 k_1 k_2 \cos \theta_{12}+2k_2^2\Big)\notag\\[5pt]
       &&\hspace{-1cm}~~
          +33 \lambda ^4 \Big[ 3 k_1^4+2 k_1 k_2 \big(2 \cos \theta_{12} (2 k_1^2+5 k_2^2)+3k_1 k_2 \cos 2\theta_{12}\big)
          +12 k_1^2 k_2^2+10 k_2^4\Big] \notag\\[5pt]
       &&\hspace{-1cm}~~
          -\lambda ^6 \Big[15 k_1^6+27k_1^4 k_2^2+135 k_1^2 k_2^4+2 k_1 k_2\Big(k_1 k_2 \big(9 \cos 2\theta_{12}(k_1^2+5 k_2^2)
          +5 k_1 k_2 \cos 3 \theta_{12}\big) \notag\\[5pt]
       &&\hspace{-1cm}~~
          \qquad +15 \cos \theta_{12} (k_1^4+3k_1^2 k_2^2+7 k_2^4)\Big)+70k_2^6\Big]\bigg\}
    \\[10pt]
    &&\hspace{-1cm}\widetilde{B}_{1h}^{4,1,s} = \frac{1}{12012 \sqrt{10}}\lambda ^4 k_2 \sin \theta_{12}
    (k_1+2 k_2 \cos \theta_{12})\bigg\{-572 \Big(8 k_1^2+8 k_1 k_2\cos \theta_{12}+7 k_2^2 \cos 2\theta_{12}+k_2^2 \Big)
    \notag\\[5pt]
    &&\hspace{-1cm}~~
       +26 \lambda ^2 \Big[80 k_1^4+5 k_1 k_2 \cos \theta_{12} (16 k_1^2+25 k_2^2)+84 k_1^2 k_2^2+\cos 2\theta_{12}
       (76 k_1^2 k_2^2+70 k_2^4)+35 k_1 k_2^3 \cos 3\theta_{12}+10 k_2^4\Big]
    \notag\\[5pt]
    &&\hspace{-1cm}~~
       -5 \lambda ^4\Big[ 112 k_1^6+66 k_1^4 k_2^2+143 k_1^2k_2^4+k_1 k_2^3 \big(\cos 3\theta_{12}
       (23k_1^2+98 k_2^2)+21 k_1 k_2 \cos 4\theta_{12}\big)
    \notag\\[5pt]
    &&\hspace{-1cm}~~
       \qquad +k_1 k_2 \cos \theta_{12} (112k_1^4+89 k_1^2 k_2^2+238 k_2^4)+2 k_2^2
       \cos 2\theta_{12} (23 k_1^4+86 k_1^2 k_2^2+49 k_2^4)+14 k_2^6 \Big] \bigg\}
  \end{eqnarray}
  \begin{eqnarray}
    &&\hspace{-1cm}\widetilde{B}_{1h}^{4,2,c} = \frac{1}{12012 \sqrt{5}}\lambda ^4 k_2^2 \sin ^2\theta_{12} \bigg\{-572 \Big(6 k_1^2+12 k_1 k_2
    \cos \theta_{12}+7 k_2^2\cos 2\theta_{12}+5 k_2^2 \Big)\notag\\[5pt]
    &&\hspace{-1cm}~~
       +26 \lambda ^2 \Big[36 k_1^4+k_1 k_2 \cos \theta_{12} (96 k_1^2+205 k_2^2)+k_2^2 \cos 2\theta_{12}
       (93 k_1^2+70 k_2^2)+123 k_1^2 k_2^2+35k_1 k_2^3 \cos 3\theta_{12}+50k_2^4 \Big]\notag\\[5pt]
    &&\hspace{-1cm}~~
       - \lambda ^4 \Big[ 180 k_1^6+303 k_1^4 k_2^2+1305 k_1^2 k_2^4+5 k_1 k_2^3 \big(\cos 3 \theta_{12}
       (38 k_1^2+98 k_2^2)+21 k_1 k_2 \cos 4\theta_{12}\big)\notag\\[5pt]
    &&\hspace{-1cm}~~
       \qquad +10 k_1 k_2 \cos \theta_{12}(36 k_1^4+101 k_1^2 k_2^2+203k_2^4)+k_2^2 \cos 2\theta_{12}
       (237k_1^4+1290 k_1^2 k_2^2+490 k_2^4)+350k_2^6 \Big] \bigg\}
    \\[10pt]
    &&\hspace{-1cm}\widetilde{B}_{1h}^{4,3,s} = \frac{1}{429 \sqrt{70}}\lambda ^4 k_2^3 \sin ^3\theta_{12} (k_1+2 k_2
       \cos \theta_{12}) \bigg\{ 286-26 \lambda ^2 \Big(2 k_1^2+5 k_1 k_2 \cos \theta_{12}+5 k_2^2 \Big) \notag\\[5pt]
    &&\hspace{-1cm}~~
       + 5 \lambda ^4 \Big[k_1^4+ k_1^3 k_2 \cos \theta_{12}+k_1^2
       k_2^2 (3 \cos 2 \theta_{12}+5)+14 k_1 k_2^3 \cos \theta_{12} +7 k_2^4 \Big] \bigg\}
  \\[10pt]
  &&\hspace{-1cm}\widetilde{B}_{1h}^{4,4,c} = \frac{1}{1716 \sqrt{35}}\lambda ^4 k_2^4 \sin ^4\theta_{12}
     \bigg\{572 -26 \lambda ^2 \Big(3 k_1^2+10 k_1 k_2 \cos \theta_{12}+10 k_2^2 \Big) \notag\\[5pt]
  &&\hspace{-1cm}~~
      + \lambda ^4 \Big[3 k_1^4+20 k_1^3 k_2 \cos \theta_{12}+30
      k_1^2 k_2^2 (\cos 2 \theta_{12}+2)+140 k_1 k_2^3 \cos \theta_{12}+70 k_2^4 \Big] \bigg\}
  \end{eqnarray}

  \begin{eqnarray}
    &&\hspace{-1cm}\widetilde{B}_{2h}^{2,1,s} =
      \frac{1}{1155 \sqrt{3} {(k_1^2+2 \cos \theta_{12} k_2 k_1+k_2^2)}^2}
      \notag\\[5pt]
      &&\hspace{-1cm}~~
      \Bigg\{2  P_m(k_1) \sin \theta_{12}(k_1+2 \cos \theta_{12} k_2) k_2{(k_1^2+2 \cos \theta_{12} k_2 k_1+k_2^2)}^2\lambda ^2\Bigg[11 (21 b^2+18 f b+5 f^2)
      \notag\\[5pt]
      &&\hspace{-1cm}~~
      -\lambda ^2 \bigg[(99 b^2+110 f b+35 f^2) k_1^2 +(99 b^2+110 f b+35 f^2) \cos \theta_{12} k_2 k_1 +(99 b^2+88 f b+25 f^2) k_2^2
      \notag\\[5pt]
      &&\hspace{-1cm}~~
      +2 f (11 b+5 f) \cos 2 \theta_{12} k_2^2\bigg]\Bigg]
      \notag\\[5pt]
      &&\hspace{-1cm}~~
      -P_m(k_2) {(k_1^2+2 \cos \theta_{12} k_2 k_1+k_2^2)}^2 \Bigg[132 f (7 b+3 f) \sin 2 \theta_{12}-11 \lambda ^2 \bigg[2 f (f \sin 4 \theta_{12} k_1+(9 b+5 f) \sin 3 \theta_{12} k_2) k_1
      \notag\\[5pt]
      &&\hspace{-1cm}~~
      +2 (21 b^2+27 f b+10 f^2) \sin \theta_{12} k_2 k_1 +\sin 2 \theta_{12} (4 f (9 b+4 f) k_1^2+3 (b+f) (7 b+5 f) k_2^2)\bigg]
      \notag\\[5pt]
      &&\hspace{-1cm}~~
      +2\sin \theta_{12} \lambda ^4 \bigg[k_1\bigg(2 (99 b^2+165 f b+70 f^2) k_2^3+9 (11 b^2+22 f b+10 f^2) k_1^2 k_2  +\cos 2 \theta_{12} (4 f (33 b+20 f) k_1^2
      \notag\\[5pt]
      &&\hspace{-1cm}~~
      +(99 b^2+220 f b+105 f^2) k_2^2) k_2+ k_1 f \Big(5 f \cos 4 \theta_{12} k_1 k_2  +\cos 3 \theta_{12} \big(66 b k_2^2+5 f (2 k_1^2+9 k_2^2)\big)\Big)\bigg)
      \notag\\[5pt]
      &&\hspace{-1cm}~~
      +\cos \theta_{12} \Big(10 f (11 b+6 f) k_1^4+27 (11 b^2+22 f b+10 f^2) k_2^2 k_1^2  +(99 b^2+165 f b+70 f^2) k_2^4\Big) \bigg] \Bigg]
      \notag\\[5pt]
      &&\hspace{-1cm}~~
      -P_m(k_3) k_2 \Bigg[ 264 f (7 b+3 f) (k_1+\cos \theta_{12} k_2) (k_1^2+2 \cos \theta_{12} k_2 k_1 +k_2^2) \sin \theta_{12}
      \notag\\[5pt]
      &&\hspace{-1cm}~~
      -11  \lambda ^2\bigg[k_2 \bigg( k_1 k_2 \Big((21 b^2+18 f b+7 f^2) \sin 4 \theta_{12} k_1 k_2+\sin 3 \theta_{12} \big(42 (k_1^2+k_2^2) b^2+54 f (k_1^2+k_2^2) b+4 f^2 (6 k_1^2+5 k_2^2)\big)\Big)
      \notag\\[5pt]
      &&\hspace{-1cm}~~
      +\sin 2 \theta_{12} \Big((21 b^2+72 f b+35 f^2) k_1^4+4 (21 b^2+36 f b+14 f^2) k_2^2 k_1^2+3 (b+f) (7 b+5 f) k_2^4\Big)\bigg)
      \notag\\[5pt]
      &&\hspace{-1cm}~~
      +2 \sin \theta_{12} k_1 \Big(2 f (9 b+5 f) k_1^4 +3 (7 b^2+21 f b+8 f^2) k_2^2 k_1^2+(21 b^2+45 f b+20 f^2) k_2^4\Big)\bigg]
      \notag\\[5pt]
      &&\hspace{-1cm}~~
      +\lambda ^4\bigg[k_1 \sin \theta_{12} \Big(10 f (11 b+7 f) k_1^6+15 f (11 b+5 f) k_2^2 k_1^4+2 (99 b^2+154 f b+50 f^2) k_2^4 k_1^2
      \notag\\[5pt]
      &&\hspace{-1cm}~~
      +(198 b^2+385 f b+175 f^2) k_2^6\Big)  + k_2 \bigg(k_1 k_2 \bigg( k_1 k_2 \Big(f (11 b+5 f) k_1 \sin 5 \theta_{12} k_2+\sin 4 \theta_{12} \big(f (11 b+10 f) k_1^2
      \notag\\[5pt]
      &&\hspace{-1cm}~~
      +(99 b^2+121 f b+45 f^2) k_2^2\big)\Big) + \sin 3 \theta_{12} \Big(5 f (11 b+9 f) k_1^4+(198 b^2+209 f b+75 f^2) k_2^2 k_1^2
      \notag\\[5pt]
      &&\hspace{-1cm}~~
      +(198 b^2+275 f b+105 f^2) k_2^4\Big) \bigg) + \sin 2 \theta_{12} \Big(15 f (11 b+7 f) k_1^6+(99 b^2+143 f b+50 f^2) k_2^2 k_1^4
      \notag\\[5pt]
      &&\hspace{-1cm}~~
      +(396 b^2+583 f b+225 f^2) k_2^4 k_1^2 +(99 b^2+165 f b+70 f^2) k_2^6\Big) \bigg)\bigg]\Bigg]\Bigg\}
    \end{eqnarray}
\newpage
\begin{eqnarray}
  &&\hspace{-1cm}\widetilde{B}_{2h}^{2,2,c} =
    \frac{ \sin ^2\theta_{12}}{1155 \sqrt{3} {(k_1^2+2 \cos \theta_{12} k_2 k_1+k_2^2)}^2}
    \notag\\[5pt]
    &&\hspace{-1cm}~~
    \Bigg\{ P_m(k_1) k_2^2 {(k_1^2+2 \cos \theta_{12} k_2 k_1+k_2^2)}^2 \lambda ^2 \Bigg[22 (21 b^2+6 f b+f^2)-\lambda ^2 \bigg[ 3 (33 b^2+22 f b+5 f^2) k_1^2
    \notag\\[5pt]
    &&\hspace{-1cm}~~
    +6 (33 b^2+22 f b+5 f^2) \cos \theta_{12} k_2 k_1+2 (99 b^2+44 f b+9 f^2) k_2^2 +4 f (11 b+3 f) \cos 2 \theta_{12} k_2^2\bigg] \Bigg]
    \notag\\[5pt]
    &&\hspace{-1cm}~~
    -P_m(k_2){(k_1^2+2 \cos \theta_{12} k_2 k_1+k_2^2)}^2\Bigg[132 f (7 b+3 f)-11 \lambda ^2 \bigg[4 f (3 b+2 f) k_1^2
    \notag\\[5pt]
    &&\hspace{-1cm}~~
    +4 f \Big(f \cos 2 \theta_{12} k_1+(9 b+5 f) \cos \theta_{12} k_2\Big) k_1+3 (b+f) (7 b+5 f) k_2^2 \bigg]  +\lambda ^4 \bigg[2 f (11 b+9 f) k_1^4
    \notag\\[5pt]
    &&\hspace{-1cm}~~
    +3 (33 b^2+88 f b+45 f^2) k_2^2 k_1^2+2 \bigg(f \Big(5 f \cos 3 \theta_{12} k_1 k_2+\cos 2 \theta_{12} \big(6 f k_1^2+66 b k_2^2+45 f k_2^2\big)\Big) k_1
    \notag\\[5pt]
    &&\hspace{-1cm}~~
    +\cos \theta_{12} k_2 \Big(3 f (22 b+15 f) k_1^2+(99 b^2+220 f b+105 f^2) k_2^2\Big)\bigg) k_1+(99 b^2+165 f b+70 f^2) k_2^4\bigg] \Bigg]
    \notag\\[5pt]
    &&\hspace{-1cm}~~
    -P_m(k_3)k_2^2  \Bigg[132 f (7 b+3 f) (k_1^2+2 \cos \theta_{12} k_2 k_1+k_2^2)-11\lambda ^2 \bigg[(21 b^2+12 f b+7 f^2) k_1^4
    \notag\\[5pt]
    &&\hspace{-1cm}~~
    +28 (3 b^2+3 f b+f^2) k_2^2 k_1^2+2 \bigg((21 b^2+18 f b+7 f^2) \cos 2 \theta_{12} k_1 k_2+2 \cos \theta_{12} \Big(3 (7 b^2+5 f b+2 f^2) k_1^2
    \notag\\[5pt]
    &&\hspace{-1cm}~~
    +(21 b^2+27 f b+10 f^2) k_2^2\Big)\bigg) k_2 k_1+3 (b+f) (7 b+5 f) k_2^4\bigg]+ \lambda ^4  \bigg[f (11 b+15 f) k_1^6
    \notag\\[5pt]
    &&\hspace{-1cm}~~
    +(99 b^2+55 f b+18 f^2) k_2^2 k_1^4+3 (132 b^2+143 f b+45 f^2) k_2^4 k_1^2+2\bigg(\Big(f (11 b+5 f) \cos 3 \theta_{12} k_1 k_2
    \notag\\[5pt]
    &&\hspace{-1cm}~~
    +\cos 2 \theta_{12} \big(f (11 b+6 f) k_1^2+(99 b^2+121 f b+45 f^2) k_2^2\big)\Big) k_1 k_2+\cos \theta_{12} \Big(15 (k_1^4+3 k_2^2 k_1^2+7 k_2^4) f^2
    \notag\\[5pt]
    &&\hspace{-1cm}~~
    +11 b (k_1^4+15 k_2^2 k_1^2+25 k_2^4) f+198 b^2 k_2^2 (k_1^2+k_2^2)\Big)\bigg) k_2 k_1+(99 b^2+165 f b+70 f^2) k_2^6\bigg] \Bigg] \Bigg\}
  \end{eqnarray}
  \begin{eqnarray}
&&\hspace{-1cm}\widetilde{B}_{2h}^{4,1,s} =
    \frac{1}{36036 \sqrt{10} {(k_1^2+2 \cos \theta_{12} k_2 k_1+k_2^2)}^2}
    \notag\\[5pt]
    &&\hspace{-1cm}~~
    \Bigg\{4 P_m(k_1) \lambda ^2 k_2  {(k_1^2+2 \cos \theta_{12} k_2 k_1+k_2^2)}^2 (k_1+2 \cos \theta_{12} k_2)\sin \theta_{12} \Bigg[208 f (11 b+5 f) -\lambda ^2 \bigg[8 (143 b^2+260 f b+105 f^2) k_1^2
    \notag\\[5pt]
    &&\hspace{-1cm}~~
    +8 (143 b^2+260 f b+105 f^2) \cos \theta_{12} k_2 k_1   +(143 b^2+1118 f b+495 f^2) k_2^2   +(1001 b^2+962 f b+345 f^2) \cos 2 \theta_{12} k_2^2\bigg]\Bigg]
    \notag\\[5pt]
    &&\hspace{-1cm}~~
    -P_m(k_2){(k_1^2+2 \cos \theta_{12} k_2 k_1+k_2^2)}^2 \Bigg[572 f^2 (2 \sin 2 \theta_{12}+7 \sin 4 \theta_{12})
    \notag\\[5pt]
    &&\hspace{-1cm}~~
    -26  \lambda ^2 f \bigg[2 (33 b+20 f) \sin \theta_{12} k_1 k_2+15 (22 b+13 f) \sin 3 \theta_{12} k_1 k_2+35 f \sin 5 \theta_{12} k_1 k_2
    \notag\\[5pt]
    &&\hspace{-1cm}~~
    +2 \sin 2 \theta_{12} \Big(2 (88 b+43 f) k_1^2+(22 b+15 f) k_2^2\Big)+\sin 4 \theta_{12} (74 f k_1^2+7 (22 b+15 f) k_2^2)\bigg]
    \notag\\[5pt]
    &&\hspace{-1cm}~~
    +\lambda ^4\bigg[315 f^2 \sin 6 \theta_{12} k_1^2 k_2^2+10 f \sin 5 \theta_{12} k_1 k_2 \big(66 f k_1^2+7 (26 b+21 f) k_2^2\big)
    \notag\\[5pt]
    &&\hspace{-1cm}~~
    +6 \sin 3 \theta_{12} k_1 \Big(6 f (169 b+100 f) k_1^2+5 (143 b^2+338 f b+168 f^2) k_2^2\Big) k_2 +2 \sin \theta_{12} k_1 k_2 \Big(2 \big(1144 b^2
    \notag\\[5pt]
    &&\hspace{-1cm}~~
    +1677 f b+675 f^2 \big) k_1^2+(429 b^2+1040 f b+525 f^2) k_2^2\Big) +\sin 2 \theta_{12} \Big(20 f (208 b+99 f) k_1^4
    \notag\\[5pt]
    &&\hspace{-1cm}~~
    +3 (2288 b^2+4472 f b+2025 f^2) k_2^2 k_1^2+2 (143 b^2+390 f b+210 f^2) k_2^4\Big)
    \notag\\[5pt]
    &&\hspace{-1cm}~~
    +\sin 4 \theta_{12} \Big(690 f^2 k_1^4+6 f (962 b+675 f) k_2^2 k_1^2+7 (143 b^2+390 f b+210 f^2) k_2^4\Big)\bigg] \Bigg]
    \notag\\[5pt]
    &&\hspace{-1cm}~~
    -P_m(k_3) k_2 \Bigg[1144 f^2 \Big(16 k_1^3+30 \cos 2 \theta_{12} k_2^2 k_1+18 k_2^2 k_1+7 \cos 3 \theta_{12} k_2^3 + \cos \theta_{12} (9 k_2^3+48 k_1^2 k_2)\Big) \sin \theta_{12}
    \notag\\[5pt]
    &&\hspace{-1cm}~~
    -26 \lambda ^2 f \bigg[\bigg(\Big(14 (11 b+5 f) \sin 5 \theta_{12} k_1 k_2^2+\sin 4 \theta_{12} k_2 \big((484 b+259 f) k_1^2+7 (22 b+15 f) k_2^2\big)
    \notag\\[5pt]
    &&\hspace{-1cm}~~
    +2 \sin 3 \theta_{12} k_1 \big((341 b+234 f) k_1^2+3 (88 b+65 f) k_2^2\big)\Big) k_2+2 \sin 2 \theta_{12} \Big(8 (44 b+35 f) k_1^4+(396 b+301 f) k_1^2 k_2^2
    \notag\\[5pt]
    &&\hspace{-1cm}~~
    +(22 b+15 f) k_2^4\Big)\bigg) k_2+2 \sin \theta_{12} k_1 \big(16 (11 b+10 f) k_1^4+(385 b+258 f) k_2^2 k_1^2+5 (11 b+8 f) k_2^4 \big)\bigg]
    \notag\\[5pt]
    &&\hspace{-1cm}~~
    +\lambda ^4 \bigg[2 k_1 \Big(80 f (26 b+21 f) k_1^6+15 f (208 b+99 f) k_2^2 k_1^4+2 (143 b^2+819 f b+675 f^2) k_2^4 k_1^2
    \notag\\[5pt]
    &&\hspace{-1cm}~~
    +(286 b^2+910 f b+525 f^2) k_2^6\Big) \sin \theta_{12}-k_2 \Bigg(k_2 \bigg(k_2 \Big(k_1 k_2 \Big(-7 (143 b^2+130 f b+45 f^2) k_1 \sin 6 \theta_{12} k_2
    \notag\\[5pt]
    &&\hspace{-1cm}~~
    -2 \sin 5 \theta_{12} \big(1001 (k_1^2+k_2^2) b^2+52 f (18 k_1^2+35 k_2^2) b+15 f^2 (22 k_1^2+49 k_2^2)\big)\Big)-\big((1001 b^2+962 f b+690 f^2) k_1^4
    \notag\\[5pt]
    &&\hspace{-1cm}~~
    +6 (715 b^2+1612 f b+675 f^2) k_2^2 k_1^2+7 (143 b^2+390 f b+210 f^2) k_2^4\big) \sin 4 \theta_{12}\Big)-2 k_1 \sin 3 \theta_{12}\big(5 f (208 b+237 f) k_1^4
    \notag\\[5pt]
    &&\hspace{-1cm}~~
    +3 (429 b^2+1378 f b+600 f^2) k_2^2 k_1^2+3 (429 b^2+1430 f b+840 f^2) k_2^4\big)\bigg)-\sin 2 \theta_{12} \Big( 240 f (26 b+21 f) k_1^6
    \notag\\[5pt]
    &&\hspace{-1cm}~~
    +2 (143 b^2+2158 f b+990 f^2) k_2^2 k_1^4+3 (715 b^2+3042 f b+2025 f^2) k_2^4 k_1^2
    \notag\\[5pt]
    &&\hspace{-1cm}~~
    +2 (143 b^2+390 f b+210 f^2) k_2^6\Big) \Bigg)\bigg]\Bigg]\Bigg\}
\end{eqnarray}
  \begin{eqnarray}
  &&\hspace{-1cm}\widetilde{B}_{2h}^{4,2,c} =
    \frac{ \sin ^2\theta_{12}}{18018 \sqrt{5} {(k_1^2+2 \cos \theta_{12} k_2 k_1+k_2^2)}^2}
    \notag\\[5pt]
    &&\hspace{-1cm}~~
    \Bigg\{2 P_m(k_1) \lambda ^2 k_2^2 {(k_1^2+2 \cos \theta_{12} k_2 k_1+k_2^2)}^2 \Bigg[104 f (11 b+3 f)-\lambda ^2 \bigg[6 (143 b^2+156 f b+45 f^2) k_1^2
    \notag\\[5pt]
    &&\hspace{-1cm}~~
    +12 (143 b^2+156 f b+45 f^2) \cos \theta_{12} k_2 k_1+(715 b^2+1066 f b+303 f^2) k_2^2+(1001 b^2+806 f b+237 f^2) \cos 2 \theta_{12} k_2^2\bigg]\Bigg]
    \notag\\[5pt]
    &&\hspace{-1cm}~~
    -P_m(k_2) {(k_1^2+2 \cos \theta_{12} k_2 k_1+k_2^2)}^2 \Bigg[572 f^2 (7 \cos 2 \theta_{12}+5)-26 \lambda ^2 f \bigg[2 (44 b+41 f) k_1^2+(264 b+205 f) \cos \theta_{12} k_2 k_1
    \notag\\[5pt]
    &&\hspace{-1cm}~~
    +35 f \cos 3 \theta_{12} k_2 k_1+5 (22 b+15 f) k_2^2+\cos 2 \theta_{12} (62 f k_1^2+7 (22 b+15 f) k_2^2)\bigg]+ \lambda ^4\bigg[6 f (104 b+101 f) k_1^4
    \notag\\[5pt]
    &&\hspace{-1cm}~~
    +3 (572 b^2+2132 f b+1305 f^2) k_2^2 k_1^2+5 f \Big(63 f \cos 4 \theta_{12} k_1 k_2+2 \cos 3 \theta_{12} \big(57 f k_1^2+7 (26 b+21 f) k_2^2\big)\Big) k_2 k_1
    \notag\\[5pt]
    &&\hspace{-1cm}~~
    +2 \cos \theta_{12} k_1 k_2 \big(3 f (624 b+505 f) k_1^2 +(1716 b^2+5330 f b+3045 f^2) k_2^2 \big) +5 (143 b^2+390 f b+210 f^2) k_2^4
    \notag\\[5pt]
    &&\hspace{-1cm}~~
    +\cos 2 \theta_{12} \big(474 f^2 k_1^4+6 f (806 b+645 f) k_2^2 k_1^2+7 (143 b^2+390 f b+210 f^2) k_2^4\big)\bigg]\Bigg]
    \notag\\[5pt]
    &&\hspace{-1cm}~~
    -P_m(k_3)k_2^2 \Bigg[572 f^2 (12 k_1^2+24 \cos \theta_{12} k_2 k_1 +7 \cos 2 \theta_{12} k_2^2+5 k_2^2)-26  \lambda ^2 f \bigg[4 (22 b+21 f) k_1^4+7 (66 b+41 f) k_2^2 k_1^2
    \notag\\[5pt]
    &&\hspace{-1cm}~~
    +14 (11 b+5 f) \cos 3 \theta_{12} k_2^3 k_1+2 \cos \theta_{12} k_1 k_2 \big(4 (55 b+36 f) k_1^2 +(319 b+205 f) k_2^2\big)+5 (22 b+15 f) k_2^4
    \notag\\[5pt]
    &&\hspace{-1cm}~~
    +\cos 2 \theta_{12} k_2^2 \big((418 b+217 f) k_1^2+7 (22 b+15 f) k_2^2\big)\bigg] + \lambda ^4 \bigg[12 f (26 b+45 f) k_1^6
    \notag\\[5pt]
    &&\hspace{-1cm}~~
    +(715 b^2+1378 f b+606 f^2) k_2^2 k_1^4+3 (1287 b^2+2782 f b+1305 f^2) k_2^4 k_1^2+\Big(7 (143 b^2+130 f b+45 f^2) \cos 4 \theta_{12} k_1 k_2
    \notag\\[5pt]
    &&\hspace{-1cm}~~
    +2 \cos 3 \theta_{12} \big(1001 (k_1^2+k_2^2) b^2 +26 f (33 k_1^2+70 k_2^2) b+15 f^2 (19 k_1^2+49 k_2^2)\big)\Big) k_2^3 k_1+2 \cos \theta_{12}k_1 k_2 \Big(12 f (26 b+45 f) k_1^4
    \notag\\[5pt]
    &&\hspace{-1cm}~~
    +(2431 b^2+4134 f b+1515 f^2) k_2^2 k_1^2+(2431 b^2+5980 f b+3045 f^2) k_2^4\Big) +5 (143 b^2+390 f b+210 f^2) k_2^6
    \notag\\[5pt]
    &&\hspace{-1cm}~~
    +\cos 2 \theta_{12} k_2^2 \Big((1001 b^2+806 f b+474 f^2) k_1^4+2 (2717 b^2+4888 f b+1935 f^2) k_2^2 k_1^2
    \notag\\[5pt]
    &&\hspace{-1cm}~~
    +7 (143 b^2+390 f b+210 f^2) k_2^4\Big)\bigg]\Bigg]\Bigg\}
  \end{eqnarray}
  \begin{eqnarray}
&&\hspace{-1cm}\widetilde{B}_{2h}^{4,3,s} =
    \frac{\sqrt{2}\sin ^3\theta_{12}}{1287\sqrt{35}{(k_1^2+2 k_2 k_1 \cos \theta_{12}+k_2^2)}^2}
    \notag\\[5pt]
  &&\hspace{-1cm}~~
    \Bigg\{P_m(k_1) k_2^3 \lambda ^4  (143 b^2+78 b f+15 f^2) (k_1+2 k_2 \cos \theta_{12}) {(k_1^2+2 k_2 k_1 \cos \theta_{12}+k_2^2)}^2
    \notag\\[5pt]
  &&\hspace{-1cm}~~
    +P_m(k_2) {(k_1^2+2 k_2 k_1 \cos \theta_{12}+k_2^2)}^2 \Bigg[572 f^2 \cos \theta_{12}-26 f \lambda ^2 \bigg[k_1 k_2 (11 b+5 f \cos 2 \theta_{12}+10 f)
    \notag\\[5pt]
  &&\hspace{-1cm}~~
    +\cos \theta_{12} \big(k_2^2 (22 b+15 f)+6 f k_1^2\big)\bigg]+ \lambda ^4 \bigg[ k_2^3 k_1 (143 b^2+10 f (26 b+21 f) \cos 2 \theta_{12} +520 b f+315 f^2)
    \notag\\[5pt]
  &&\hspace{-1cm}~~
    +k_2^4 (143 b^2+390 b f+210 f^2) \cos \theta_{12}+6 f k_2 k_1^3 (13 b+10 f \cos 2 \theta_{12}+15 f)
    \notag\\[5pt]
  &&\hspace{-1cm}~~
    +18 f k_2^2 k_1^2 \cos \theta_{12} (26 b+5 f \cos 2 \theta_{12}+20 f)+30 f^2 k_1^4 \cos \theta_{12}\bigg]\Bigg]
    \notag\\[5pt]
  &&\hspace{-1cm}~~
    +P_m(k_3) k_2^3 \Bigg[572 f^2 (k_2 \cos \theta_{12}+k_1) -26 f \lambda ^2 \bigg[k_2 k_1^2 (44 b+21 f) \cos \theta_{12}+k_2^2 k_1 \big(2 (11 b+5 f) \cos 2 \theta_{12}+33 b+20 f \big)
    \notag\\[5pt]
  &&\hspace{-1cm}~~
    +k_2^3 (22 b+15 f) \cos \theta_{12}+k_1^3 (11 b+6 f)\bigg] +\lambda ^4 \bigg[k_2 k_1^4 (143 b^2+78 b f+30 f^2) \cos \theta_{12}+2 k_2^2 k_1^3 \Big((143 b^2+104 b f
    \notag\\[5pt]
  &&\hspace{-1cm}~~
    +30 f^2) \cos 2 \theta_{12} +143 b^2+169 b f+45 f^2 \Big)+2 k_2^3 k_1^2 \cos \theta_{12} \Big((143 b^2+130 b f+45 f^2) \cos 2 \theta_{12}
    \notag\\[5pt]
  &&\hspace{-1cm}~~
    +286 b^2+494 b f+180 f^2\Big)+k_2^4 k_1 \Big((286 b^2+520 b f+210 f^2) \cos 2 \theta_{12} +286 b^2+650 b f+315 f^2\Big)
    \notag\\[5pt]
  &&\hspace{-1cm}~~
    +k_2^5 (143 b^2+390 b f+210 f^2) \cos \theta_{12}+15 f^2 k_1^5 \bigg]\Bigg] \Bigg\}
  \\[10pt]
&&\hspace{-1cm}\widetilde{B}_{2h}^{4,4,c} =
    \frac{\sin ^4\theta_{12}}{2574 \sqrt{35} {(k_1^2+2 k_2 k_1 \cos \theta_{12}+k_2^2)}^2}
    \notag\\[5pt]
  &&\hspace{-1cm}~~
    \Bigg\{ 2 P_m(k_1) \lambda ^4 k_2^4 (143 b^2+26 b f+3 f^2) {(k_1^2+2 k_2 k_1 \cos \theta_{12}+k_2^2)}^2
    \notag\\[5pt]
  &&\hspace{-1cm}~~
    +P_m(k_2) {(k_1^2+2 k_2 k_1 \cos \theta_{12}+k_2^2)}^2 \Bigg[572 f^2 -26 f \lambda ^2 \bigg[k_2^2 (22 b+15 f)+10 f k_2 k_1 \cos \theta_{12}+2 f k_1^2 \bigg]
    \notag\\[5pt]
  &&\hspace{-1cm}~~
     + \lambda ^4 \bigg[k_2^4 (143 b^2+390 b f+210 f^2)+6 f k_2^2 k_1^2 (26 b+15 f \cos 2 \theta_{12}+30 f)+20 f k_2^3 k_1 (26 b+21 f) \cos \theta_{12}
     \notag\\[5pt]
  &&\hspace{-1cm}~~
     +60 f^2 k_2 k_1^3 \cos \theta_{12}+6 f^2 k_1^4 \bigg]\Bigg]
     \notag\\[5pt]
  &&\hspace{-1cm}~~
    +P_m(k_3) k_2^4 \Bigg[572 f^2 -26 f \lambda ^2 \bigg[k_1^2 (22 b+7 f)+4 k_2 k_1 (11 b+5 f) \cos \theta_{12}+k_2^2 (22 b+15 f)\bigg]
    \notag\\[5pt]
  &&\hspace{-1cm}~~
    + \lambda ^4 \bigg[4 k_2 k_1^3 (143 b^2+78 b f+15 f^2) \cos \theta_{12} +2 k_2^2 k_1^2 \Big((143 b^2+130 b f+45 f^2) \cos 2 \theta_{12}
    \notag\\[5pt]
  &&\hspace{-1cm}~~
    +286 b^2+338 b f+90 f^2\Big)+4 k_2^3 k_1 (143 b^2+260 b f+105 f^2) \cos \theta_{12}+k_1^4 (143 b^2+26 b f+6 f^2)
    \notag\\[5pt]
  &&\hspace{-1cm}~~
    +k_2^4 (143 b^2+390 b f+210 f^2)\bigg] \Bigg] \Bigg\}
  \end{eqnarray}
  \begin{eqnarray}
&&\hspace{-1cm}\widetilde{B}_{3h}^{2,1,s}=\frac{f}{16170 \sqrt{3} k_1 k_2 {(k_1^2+2 \cos \theta_{12} k_2 k_1+k_2^2)}^2}
    \notag\\[5pt]
    &&\hspace{-1cm}~~
    \Bigg\{-P_m(k_1) P_m(k_2) (k_1^2+2 \cos \theta_{12} k_2 k_1+k_2^2) \bigg[f^2 (35 f+44) \sin 6 \theta_{12} k_1 k_2^3+\sin 5 \theta_{12} \Big(\big(462 b^2+66 f (7 f+9) b
    \notag\\[5pt]
    &&\hspace{-1cm}~~
    +f^2 (315 f+341)\big) k_1^2+7 f^2 (5 f+11) k_2^2\Big) k_2^2 +\sin 4 \theta_{12} k_1 k_2 \Big(\big(693 (f+3) b^2+198 f (7 f+10) b +f^2 (595 f+682)\big) k_1^2
    \notag\\[5pt]
    &&\hspace{-1cm}~~
    +\big(231 (6 f+13) b^2+308 f (7 f+9) b+2 f^2 (420 f+517)\big) k_2^2\Big) +\sin 3 \theta_{12} \bigg(7 \Big(33 (3 f+7) b^2 +66 f (2 f+3) b
    \notag\\[5pt]
    &&\hspace{-1cm}~~
    +5 f^2 (9 f+11)\Big) k_1^4 +\Big(2520 f^3+154 (59 b+19) f^2+99 (3 b (35 b+36)+14 b_2) f +231 b (b (14 b+55)+14 b_2)\Big) k_2^2 k_1^2
    \notag\\[5pt]
    &&\hspace{-1cm}~~
    +7 \Big(75 f^3+11 (22 b+9) f^2+198 b (b+2) f+462 b^2\Big) k_2^4\bigg) +\sin 2 \theta_{12} k_1 k_2 \bigg(2 \Big(1120 f^3+11 (434 b+113) f^2
    \notag\\[5pt]
    &&\hspace{-1cm}~~
    +99 (10 b (7 b+5)+7 b_2) f+231 b \big(2 b (7 b+13)+7 b_2\big)\Big) k_1^2 +\Big(2625 f^3+22 (504 b+145) f^2
    \notag\\[5pt]
    &&\hspace{-1cm}~~
    +1386 (b (11 b+10)+b_2) f +462 b \big(2 b (7 b+18)+7 b_2\big)\Big) k_2^2\bigg) +\sin \theta_{12} \bigg(7 \Big(75 f^3+11 (34 b+5) f^2
    \notag\\[5pt]
    &&\hspace{-1cm}~~
     +99 b (7 b+2) f+231 b^2 (2 b+1)\Big) k_1^4+\Big(2625 f^3+11 (1092 b+307) f^2 +99 \big(3 b (63 b+50)+14 b_2\big) f
    \notag\\[5pt]
    &&\hspace{-1cm}~~
    +1617 b \big(b (6 b+11)+2 b_2\big)\Big) k_2^2 k_1^2 +14 \Big(50 f^3+22 (11 b+2) f^2+198 b (2 b+1) f+231 b^2 (b+1)\Big) k_2^4\bigg)\bigg]
    \notag\\[5pt]
    &&\hspace{-1cm}~~
    +2 P_m(k_2) P_m(k_3) k_1 (k_1+2 \cos \theta_{12} k_2) \sin \theta_{12} \bigg[7 \Big(60 f^3+11 (23 b+5) f^2+198 b (2 b+1) f+231 b^2 (b+1)\Big) k_1^4
    \notag\\[5pt]
    &&\hspace{-1cm}~~
    +\Big(525 f^3+77 (23 b+5) f^2+99 (3 b (7 b+4)-28 b_2) f +231 b \big(b (7 b+4)-28 b_2\big)\Big) k_2^2 k_1^2 +\bigg(33 f^2 \cos 4 \theta_{12} k_2^2
    \notag\\[5pt]
    &&\hspace{-1cm}~~
    +\Big(231 (3 f+5) b^2+66 f (7 f+12) b+f^2 (105 f+242)\Big) \cos 3 \theta_{12} k_1 k_2 +\cos 2 \theta_{12} \Big(7 \Big(33 (3 f+7) b^2
    \notag\\[5pt]
    &&\hspace{-1cm}~~
    +66 f (2 f+3) b+5 f^2 (9 f+11)\Big) k_1^2+2 \Big(105 f^3+22 (21 b+8) f^2+99 \big(b (7 b+8)-7 b_2\big) f
    \notag\\[5pt]
    &&\hspace{-1cm}~~
    +231 b (5 b-7 b_2)\Big) k_2^2\Big)\bigg) k_1^2 +\cos \theta_{12} k_1 k_2 \bigg(\Big(1365 f^3+22 (224 b+59) f^2+99 (63 b^2+48 b-14 b_2) f
    \notag\\[5pt]
    &&\hspace{-1cm}~~
    +231 b \big(b (14 b+23)-14 b_2\big)\Big) k_1^2-1386 b_2 (7 b+3 f) k_2^2\bigg) -462 b_2 (7 b+3 f) k_2^4\bigg]
    \notag\\[5pt]
    &&\hspace{-1cm}~~
    +P_m(k_3) P_m(k_1)  k_2^2\bigg[\bigg(33 f^2 \sin 6 \theta_{12} k_1 k_2^2+\sin 5 \theta_{12} \Big(99 (7 b^2+6 f b+2 f^2) k_1^2+7 f^2 (5 f+11) k_2^2\Big) k_2
    \notag\\[5pt]
    &&\hspace{-1cm}~~
    +\sin 3 \theta_{12} k_2 \bigg(\Big(945 f^3 +77 (42 b+17) f^2+99 (b (35 b+52)-14 b_2) f+3234 b (2 b-b_2)\Big) k_1^2 +7 \Big(75 f^3+11 (22 b+9) f^2
    \notag\\[5pt]
    &&\hspace{-1cm}~~
    +198 b (b+2) f +462 b^2\Big) k_2^2 \bigg) +\sin 2 \theta_{12} k_1 \bigg(2 \Big(245 f^3+110 (7 b+2) f^2 +99 (b (7 b+8)-21 b_2) f+231 b (4 b-21 b_2)\Big) k_1^2
    \notag\\[5pt]
    &&\hspace{-1cm}~~
    +\Big(2100 f^3+11 (686 b+193) f^2+1386 (6 b (b+1)-b_2) f +462 b (b (7 b+20)-7 b_2)\Big) k_2^2\bigg) +\sin 4 \theta_{12} k_1 \Big(33 (21 b^2
    \notag\\[5pt]
    &&\hspace{-1cm}~~
    +18 f b+5 f^2) k_1^2+\Big(693 (2 f+5) b^2+308 f (4 f+9) b +2 f^2 (210 f+407)\Big) k_2^2\Big)\bigg) k_2+\sin \theta_{12} \bigg(-924 (7 b
    \notag\\[5pt]
    &&\hspace{-1cm}~~
    +3 f) b_2 k_1^4+\Big(1575 f^3+11 (448 b+113) f^2 +99 \big(b (49 b+38)-42 b_2\big) f+231 b \big(b (14 b+13)-42 b_2\big)\Big) k_2^2 k_1^2
    \notag\\[5pt]
    &&\hspace{-1cm}~~
    +14 \Big(50 f^3+22 (11 b+2) f^2+198 b (2 b+1) f+231 b^2 (b+1)\Big) k_2^4\bigg)\bigg] \Bigg\}
  \end{eqnarray}
  \begin{eqnarray}
&&\hspace{-1cm}\widetilde{B}_{3h}^{2,2,c}=\frac{ f \sin ^2\theta_{12}}{8085 \sqrt{3} k_1 k_2 {(k_1^2+2 \cos \theta_{12} k_2 k_1+k_2^2)}^2}
    \notag\\[5pt]
    &&\hspace{-1cm}~~
    \Bigg\{-P_m(k_1) P_m(k_2) (k_1^2+2 \cos \theta_{12} k_2 k_1+k_2^2) \Bigg[ k_2 \Bigg(k_1^3\Big(357 f^3
    \notag\\[5pt]
    &&\hspace{-1cm}~~
    +11 (147 b+31) f^2 +33 \big(b (91 b+50)+7 b_2\big) f +231 b \big(b (7 b+19)+7 b_2 \big)\Big) + k_2^2 k_1 \Big(567 f^3+11 (217 b+65) f^2
    \notag\\[5pt]
    &&\hspace{-1cm}~~
    +33 \big(2 b (56 b+55)+7 b_2\big) f +231 b \big(b (7 b+29)+7 b_2\big)\Big) +f^2 (35 f+44) \cos 4 \theta_{12} k_2^2 k_1
    +\cos 2 \theta_{12}  k_1 \bigg(\Big(693 (f+3) b^2
    \notag\\[5pt]
    &&\hspace{-1cm}~~
    +66 f (14 f+17) b+f^2 (273 f+275)\Big) k_1^2
    +\Big(231 (6 f+13) b^2+22 f (77 f+87) b+f^2 (518 f+627)\Big) k_2^2\bigg)
    \notag\\[5pt]
    &&\hspace{-1cm}~~
    +\cos 3 \theta_{12} k_2 \Big(\big(462 b^2+462 f (f+1) b
    +f^2 (203 f+209)\big) k_1^2+7 f^2 (5 f+11) k_2^2\Big)\Bigg)
    +\cos \theta_{12} \bigg(7 \Big(33 (3 f+7) b^2
    \notag\\[5pt]
    &&\hspace{-1cm}~~
    +66 f (f+1) b+f^2 (15 f+11)\Big) k_1^4
    +\Big(1092 f^3+22 (203 b+57) f^2+231 \big(b (29 b+24)+2 b_2\big) f+231 b (b (14 b+47)
    \notag\\[5pt]
    &&\hspace{-1cm}~~
    +14 b_2)\Big) k_2^2 k_1^2
    +7 \Big(66 (3 f+7) b^2+88 f (2 f+3) b+5 f^2 (9 f+11)\Big) k_2^4\bigg)\Bigg]
    \notag\\[5pt]
    &&\hspace{-1cm}~~
    +P_m(k_2) P_m(k_3)k_1 \Bigg[-462 b_2 (7 b+3 f) k_2^5+\Big(315 f^3+11 (98 b+19) f^2+33 (5 b (7 b+4)-91 b_2) f
    \notag\\[5pt]
    &&\hspace{-1cm}~~
    +231 b (4 b-35 b_2)\Big) k_1^2 k_2^3+\Big(378 f^3+55 (28 b+5) f^2+33 (56 b^2+34 b-7 b_2) f+231 b (9 b-7 b_2)\Big) k_1^4 k_2
    \notag\\[5pt]
    &&\hspace{-1cm}~~
    +k_1 \Bigg( k_1 k_2 \bigg(33 f^2 \cos 4 \theta_{12} k_2^2+\cos 3 \theta_{12} k_1 k_2 \Big(231 (3 f+5) b^2+66 f (7 f+9) b+f^2 (105 f+176)\Big)
    \notag\\[5pt]
    &&\hspace{-1cm}~~
    +2 \cos 2 \theta_{12} \Big(231 \big(3 (f+2) k_1^2+(3 f+5) k_2^2\big) b^2+33 \big(2 f (7 f+9) (k_1^2+k_2^2)-49 b_2 k_2^2\big) b
    \notag\\[5pt]
    &&\hspace{-1cm}~~
    +f \big(f (21 f+22) (6 k_1^2+5 k_2^2)-693 b_2 k_2^2\big)\Big)\bigg)+\cos \theta_{12} \Big(7 \Big(33 (3 f+7) b^2+66 f (f+1) b+f^2 (15 f+11)\Big) k_1^4
    \notag\\[5pt]
    &&\hspace{-1cm}~~
    +\Big(945 f^3+22 (161 b+34) f^2+33 \big(b (133 b+94)-56 b_2\big) f+231 b (23 b-28 b_2)\Big) k_2^2 k_1^2 -1386 b_2 (7 b+3 f) k_2^4\Big)\Bigg) \Bigg]
    \notag\\[5pt]
    &&\hspace{-1cm}~~
    +P_m(k_3) P_m(k_1) k_2^3\Bigg[ k_1^3 \Big(35 f^3+11 (7 b+4) f^2-33 \big(b (7 b-8)+7 b_2\big) f -231 b \big(b (7 b-4)+7 b_2\big)\Big)
    \notag\\[5pt]
    &&\hspace{-1cm}~~
    +k_2^2 k_1\Big(378 f^3+11 (119 b+40) f^2+ 33 \big(2 b (7 b+29)-7 b_2\big) f-231 b \big(b (7 b-13)+7 b_2\big)\Big)
    \notag\\[5pt]
    &&\hspace{-1cm}~~
    +33 f^2 \cos 4 \theta_{12} k_2^2 k_1+\cos 2 \theta_{12}k_1 \Big(33 k_1^2 (21 b^2+6 f b+f^2)+ k_2^2 \big(693 (2 f+5) b^2+22 f (35 f+81) b
    \notag\\[5pt]
    &&\hspace{-1cm}~~
    +f^2 (252 f+451)\big) \Big) +\cos 3 \theta_{12} k_2 \Big(99 (7 b^2+4 f b+f^2) k_1^2+7 f^2 (5 f+11) k_2^2\Big)
    \notag\\[5pt]
    &&\hspace{-1cm}~~
    +\cos \theta_{12} k_2 \bigg(k_1^2 \Big(315 f^3+44 (21 b+10) f^2+33 \big(b (7 b+58)-14 b_2\big) f-462 b \big(b (7 b-9)+7 b_2\big)\Big)
    \notag\\[5pt]
    &&\hspace{-1cm}~~
    +7 k_2^2 \Big(66 (3 f+7) b^2+88 f (2 f+3) b+5 f^2 (9 f+11)\Big) \bigg)\Bigg]
    \Bigg\}
  \end{eqnarray}
  \begin{eqnarray}
    &&\hspace{-1cm}\widetilde{B}_{3h}^{4,1,s}=\frac{f^2}{252252 \sqrt{10} k_1 k_2 {(k_1^2+2 \cos \theta_{12} k_2 k_1+k_2^2)}^2}
    \notag\\[5pt]
    &&\hspace{-1cm}~~
    \Bigg\{
    -P_m(k_1) P_m(k_2) (k_1^2+2 \cos \theta_{12} k_2 k_1+k_2^2) \Biggl[\big(91 b (35 f+44)+2 f (1155 f+962)\big) \sin 6 \theta_{12} k_1 k_2^3
    \notag\\[5pt]
    &&\hspace{-1cm}~~
    +\sin 5 \theta_{12} k_2^2 \bigg(\Big(11970 f^2+13 (1799 b+727) f+143 b (98 b+125)\Big) k_1^2+7 \Big(91 b (5 f+11)+f (330 f+481)\Big) k_2^2\bigg)
    \notag\\[5pt]
    &&\hspace{-1cm}~~
    +\sin 4 \theta_{12} k_1 k_2 \bigg(\Big(29029 b^2+247 (168 f+121) b+f (17955 f+14807)\Big) k_1^2+13 \Big(1890 f^2+(4802 b+1903) f
    \notag\\[5pt]
    &&\hspace{-1cm}~~
    +77 b (44 b+51)\Big) k_2^2\bigg) +\sin 2 \theta_{12} k_1 k_2 \bigg(26 \Big(4543 b^2+(5768 f+3861) b+f (1785 f+1661)+616 b_2\Big) k_1^2
    \notag\\[5pt]
    &&\hspace{-1cm}~~
    +\Big(49770 f^2+13 (12061 b+3590) f+2002 (60 b^2+55 b+8 b_2)\Big) k_2^2\bigg) +\sin \theta_{12} \bigg(7 \Big(5005 b^2 +286 (19 f+8) b +5 f (297 f
    \notag\\[5pt]
    &&\hspace{-1cm}~~
    +208)\Big) k_1^4+\Big(47565 f^2+26 (5992 b+1847) f+143 \big(b (875 b+794)+112 b_2\big)\Big) k_2^2 k_1^2 +14 \Big(675 f^2 +13 (149 b+43) f
    \notag\\[5pt]
    &&\hspace{-1cm}~~
    +143 b (11 b+9)\Big) k_2^4\bigg)+\sin 3 \theta_{12} \bigg(7 \Big(2145 b^2+26 (117 f+88) b+5 f (237 f+208)\Big) k_1^4 +\Big(58275 f^2+91 (1881 b+611) f
    \notag\\[5pt]
    &&\hspace{-1cm}~~
    +143 \big(b (889 b+871)+112 b_2\big)\Big) k_2^2 k_1^2+7 \Big(1800 f^2+39 (143 b+41) f +715 b (6 b+5)\Big) k_2^4\bigg)\Biggr]
    \notag\\[5pt]
    &&\hspace{-1cm}~~
    +2  P_m(k_2) P_m(k_3) k_1 (k_1+2 \cos \theta_{12} k_2)\sin \theta_{12}  \Biggl[7 \Big(3575 b^2+26 (163 f+88) b+5 f (267 f+208)\Big) k_1^4+\Big(7455 f^2
    \notag\\[5pt]
    &&\hspace{-1cm}~~
    +1183 (17 b+5) f+143 \big(b (70 b+103)-126 b_2\big)\Big) k_2^2 k_1^2+ k_1 k_2 \bigg( \cos 4 \theta_{12} k_1 k_2 \Big(455 b (7 f+11)+3 f (245 f+481)\Big)
    \notag\\[5pt]
    &&\hspace{-1cm}~~
    +\cos 3 \theta_{12} \Big(\big(6195 f^2+13 (1554 b+499) f+143 b (105 b+113)\big) k_1^2-14014 b_2 k_2^2\Big)\bigg)+\cos \theta_{12} k_1 k_2 \bigg(\Big(29085 f^2
    \notag\\[5pt]
    &&\hspace{-1cm}~~
    +13 (6286 b+1741) f +715 b (91 b+67)-16016 b_2\Big) k_1^2-34034 b_2 k_2^2\bigg)-2002 b_2 k_2^4+\cos 2 \theta_{12} \bigg(7 k_1^4 \Big(2145 b^2+26 (117 f
    \notag\\[5pt]
    &&\hspace{-1cm}~~
    +88) b +5 f (237 f+208)\Big) +2 k_2^2 k_1^2 \Big(4725 f^2+13 (1064 b+277) f +143 \big(b (105 b+43)-105 b_2\big)\Big) -14014 b_2 k_2^4\bigg)\Biggr]
    \notag\\[5pt]
    &&\hspace{-1cm}~~
    + P_m(k_3) P_m(k_1) k_2^2\Biggl[k_2 \Bigg(39 (77 b+37 f) \sin 6 \theta_{12} k_1 k_2^2+\sin 3 \theta_{12} k_2 \bigg( k_1^2 \Big(24885 f^2+91 (714 b+269) f
    \notag\\[5pt]
    &&\hspace{-1cm}~~
    +143 \big(b (231 b+409)-112 b_2\big)\Big) +7 k_2^2 \Big(1800 f^2+39 (143 b+41) f+715 b (6 b+5)\Big) \bigg)
    \notag\\[5pt]
    &&\hspace{-1cm}~~
    +\sin 5 \theta_{12} k_2 \Big(39 (253 b+117 f) k_1^2+7 \big(91 b (5 f+11)+f (330 f+481)\big) k_2^2\Big)
    \notag\\[5pt]
    &&\hspace{-1cm}~~
    +\sin 2 \theta_{12} k_1 \bigg(16 \Big(735 f^2+260 (7 b+2) f+143 \big(b (7 b+8)-21 b_2\big)\Big) k_1^2+\Big(41580 f^2+13 (9212 b+2521) f
    \notag\\[5pt]
    &&\hspace{-1cm}~~
    +1001 \big(96 b^2+69 b-16 b_2\big)\Big) k_2^2\bigg)+\sin 4 \theta_{12} k_1 \Big(624 (11 b+5 f) k_1^2+\big(16016 b^2
    \notag\\[5pt]
    &&\hspace{-1cm}~~
    +91 (382 f+451) b+f (14490 f+17849)\big) k_2^2\Big)\Bigg) +\sin \theta_{12} \bigg(-32032 b_2 k_1^4+\Big(31185 f^2+26 (3143 b+778) f
    \notag\\[5pt]
    &&\hspace{-1cm}~~
    +1573 b (49 b+20)-48048 b_2\Big) k_2^2 k_1^2+14 \Big(675 f^2+13 (149 b+43) f+143 b (11 b+9)\Big) k_2^4\bigg)\Biggr]
    \Bigg\}
  \end{eqnarray}
  \begin{eqnarray}
&&\hspace{-1cm}\widetilde{B}_{3h}^{4,2,c}=\frac{f^2\sin ^2\theta_{12}}{126126 \sqrt{5} k_1 k_2 {(k_1^2+2 \cos \theta_{12} k_2 k_1+k_2^2)}^2}
    \notag\\[5pt]
    &&\hspace{-1cm}~~
      \Bigg\{ -P_m(k_1) P_m(k_2) (k_1^2+2 \cos \theta_{12} k_2 k_1+k_2^2) \Bigg[k_2 \Bigg(2  k_1^3 \Big(6132 f^2+39 (539 b+124) f+143 \big(b (133 b+100)+14 b_2\big)\Big)
     \notag\\[5pt]
     &&\hspace{-1cm}~~
    +k_2^2 k_1 \Big(19089 f^2+13 (4669 b+1448) f+286 (b (175 b+164)+14 b_2)\Big) + \cos 4 \theta_{12} k_2^2 k_1 \Big(91 b (35 f+44)
    \notag\\[5pt]
    &&\hspace{-1cm}~~
    +f (1995 f+1612)\Big)+2 k_1\cos 2 \theta_{12} \bigg(\Big(5208 f^2 +195 (77 b+20) f+143 b (91 b+68)\Big) k_1^2
    \notag\\[5pt]
    &&\hspace{-1cm}~~
    +\Big(9618 f^2+39 (665 b+242) f+143 b (133 b+158)\Big) k_2^2\bigg)  +\cos 3 \theta_{12} k_2 \bigg(\Big(14014 b^2
    \notag\\[5pt]
    &&\hspace{-1cm}~~
    +91 (221 f+165) b+f (8631 f+6565)\Big) k_1^2+7 \Big(91 b (5 f+11)+f (285 f+403)\Big) k_2^2\bigg)\Bigg)
    \notag\\[5pt]
    &&\hspace{-1cm}~~
    +\cos \theta_{12} \bigg(28 \Big(135 f^2+78 (6 b+1) f+143 b (3 b+2)\Big) k_1^4+\Big(37989 f^2+13 (9205 b+2687) f
    \notag\\[5pt]
    &&\hspace{-1cm}~~
    +1001 \big(b (102 b+89)+8 b_2\big)\Big) k_2^2 k_1^2+7 \Big(1515 f^2+13 (349 b+113) f+143 b (24 b+25)\Big) k_2^4\bigg)\Bigg]
    \notag\\[5pt]
    &&\hspace{-1cm}~~
    +P_m(k_2) P_m(k_3)k_1 \Bigg[ k_2\Bigg(2 \Big(6363 f^2+39 (511 b+100) f+286 \big(56 b^2+34 b-7 b_2\big)\Big) k_1^4+\Big(9135 f^2+13 (2009 b+421) f
    \notag\\[5pt]
    &&\hspace{-1cm}~~
    +143 \big(b (140 b+87)-266 b_2\big)\Big) k_2^2 k_1^2+ k_1 k_2 \bigg( k_1 k_2\cos 4 \theta_{12}\big(455 b (7 f+11)+3 f (245 f+403)\big)
    \notag\\[5pt]
    &&\hspace{-1cm}~~
    +\cos 3 \theta_{12} \Big(\big(5985 f^2+13 (1547 b+433) f+143 b (84 b+121)\big) k_1^2-14014 b_2 k_2^2\Big)\bigg)
    \notag\\[5pt]
    &&\hspace{-1cm}~~
    -10010 b_2 k_2^4+2 \cos 2 \theta_{12} \bigg(3  k_1^4 \Big(1659 f^2+13 (385 b+96) f+572 b (7 b+6)\Big)+  k_2^2 k_1^2 \Big(4515 f^2+13741 b f
    \notag\\[5pt]
    &&\hspace{-1cm}~~
    +429 b (28 b+17)+247 (13 f-77 b_2)\Big)-7007 b_2 k_2^4\bigg)\Bigg)+\cos \theta_{12} k_1 \bigg(28 k_1^4 \Big(135 f^2+78 (6 b+1) f
    \notag\\[5pt]
    &&\hspace{-1cm}~~
    +143 b (3 b+2)\Big) +k_2^2 k_1^2 \Big(31815 f^2+13 (7189 b+1583) f+143 \big(b (532 b+327)-224 b_2\big)\Big) -58058 b_2 k_2^4\bigg)\Bigg]
    \notag\\[5pt]
    &&\hspace{-1cm}~~
    +P_m(k_3) P_m(k_1) k_2^3\Bigg[4 \Big(315 f^2+78 (7 b+4) f-143 (b (7 b-8)+7 b_2)\Big) k_1^3 +\Big(12726 f^2
    \notag\\[5pt]
    &&\hspace{-1cm}~~
    +65 (532 b+185) f+143 (b (154 b+197)-28 b_2)\Big) k_2^2 k_1+39 (77 b+31 f) \cos 4 \theta_{12} k_2^2 k_1
    \notag\\[5pt]
    &&\hspace{-1cm}~~
    +2 \cos 2 \theta_{12} k_1 \Big(156 (11 b+3 f) k_1^2+\big(4977 f^2+13 (938 b+499) f+715 b (7 b+23)\big) k_2^2\Big)
    \notag\\[5pt]
    &&\hspace{-1cm}~~
    +\cos \theta_{12} k_2 \bigg(4 \Big(2835 f^2+312 (21 b+10) f+143 \big(b (7 b+58)-14 b_2\big)\Big) k_1^2+7 \Big(1515 f^2+13 (349 b+113) f
    \notag\\[5pt]
    &&\hspace{-1cm}~~
    +143 b (24 b+25)\Big) k_2^2\bigg)+\cos 3 \theta_{12} k_2 \Big(312 (22 b+9 f) k_1^2+7 \big(91 b (5 f+11)+f (285 f+403)\big) k_2^2\Big)\Bigg]
    \Bigg\}
  \end{eqnarray}
  \newpage
\begin{eqnarray}
  &&\hspace{-1cm}\widetilde{B}_{3h}^{4,3,s}=\frac{f^2 \sin ^3\theta_{12}} {9009 \sqrt{70} k_1 k_2 {(k_1^2+2 \cos \theta_{12} k_2 k_1+k_2^2)}^2}
    \notag\\[5pt]
    &&\hspace{-1cm}~~
    \Bigg\{P_m(k_1) P_m(k_2) (k_1^2+2 k_2 k_1 \cos \theta_{12}+k_2^2) \Biggr[7 k_1^4 (143 b^2+78 b f +15 f^2)
    +2 k_2 \bigg(k_2 \Big(\cos 2 \theta_{12} \Big(k_1^2 \big(13 (161 b+33) f
    \notag\\[5pt]
    &&\hspace{-1cm}~~
    +143 b (14 b+11)+630 f^2\big) +7 k_2^2 \big(13 b (5 f+11)+3 f (10 f+13)\big)\Big)+k_1 k_2 \cos 3 \theta_{12}\big(13 b (35 f+44)
    \notag\\[5pt]
    &&\hspace{-1cm}~~
    +6 f (35 f+26)\big)\Big) +k_1 \cos \theta_{12} \Big(7 k_1^2 \big(39 (8 b+1) f+143 b (3 b+1)+75 f^2\big)
    \notag\\[5pt]
    &&\hspace{-1cm}~~
    +k_2^2 \big(13 (343 b+93) f+143 b (28 b+31)+1260 f^2\big)\Big)\bigg)+k_2^2 k_1^2 \Big(442 (14 b+3) f+143 b (49 b+34)+1575 f^2\Big)
    \notag\\[5pt]
    &&\hspace{-1cm}~~
    +14 k_2^4 \Big(13 (13 b+3) f+143 b (b+1) +45 f^2\Big)\Biggr]
    \notag\\[5pt]
    &&\hspace{-1cm}~~
    - P_m(k_2) P_m(k_3) k_1 (k_1+ 2 k_2 \cos \theta_{12}) \Biggr[7 k_1^4 (143 b^2+78 b f+15 f^2) +2 k_2 k_1 \bigg(7 \cos \theta_{12} \Big(k_1^2 \big(39 (6 b+1) f
    \notag\\[5pt]
    &&\hspace{-1cm}~~
    +143 b (b+1)+75 f^2 \big)-286 b_2 k_2^2\Big)+k_1 k_2  \cos 2 \theta_{12} \big(65 b (7 f+11) +3 f (35 f+39)\big)\bigg)
    \notag\\[5pt]
    &&\hspace{-1cm}~~
    +2 k_2^2 k_1^2 \Big(13 (91 b+12) f+143 \big(b (7 b+2)-7 b_2\big)+420 f^2\Big)-2002 b_2 k_2^4 \Biggr]
    \notag\\[5pt]
    &&\hspace{-1cm}~~
    - P_m(k_3) P_m(k_1)k_2^4 \Biggr[k_1^2 \Big(78 (7 b+4) f-143 b (7 b-8)+315 f^2\Big)+14 k_2^2 \Big(13 (13 b+3) f+143 b (b+1)
    \notag\\[5pt]
    &&\hspace{-1cm}~~
    +45 f^2\Big)  +36\cos \theta_{12} k_1  k_2  \Big(13 b (7 f+11)+f (35 f+39)\Big) +78 \cos 3 \theta_{12} k_2 k_1 (11 b+3 f)
    \notag\\[5pt]
    &&\hspace{-1cm}~~
    +\cos 2 \theta_{12} \Big(78 k_1^2 (11 b+3 f)+14 k_2^2 \big(13 b (5 f+11)+3 f (10 f+13)\big)\Big)\Biggr]\Bigg\}
    \\[10pt]
  &&\hspace{-1cm}\widetilde{B}_{3h}^{4,4,c}=\frac{f^2 \sin ^4\theta_{12}}{9009 \sqrt{35} k_1 {(k_1^2+2 \cos \theta_{12} k_2 k_1+k_2^2)}^2}
        \notag\\[5pt]
        &&\hspace{-1cm}~~
        \Bigg\{ P_m(k_1) P_m(k_2) (k_1^2+2 k_2 k_1 \cos \theta_{12}+k_2^2) \Bigg[7 k_1^3 (143 b^2+39 b f+6 f^2)+k_2 \cos \theta_{12} \bigg(7 k_1^2 \Big(286 b^2+143 b (f+1)
        \notag\\[5pt]
        &&\hspace{-1cm}~~
        +f (27 f+13)\Big)+2 k_1 k_2 \cos \theta_{12} \Big(13 b (35 f+44)+f (105 f+52)\Big)+7 k_2^2 \Big(13 b (5 f+11)+f (15 f+13)\Big)\bigg)
        \notag\\[5pt]
        &&\hspace{-1cm}~~
        +k_2^2 k_1 \Big(39 (7 b+2) f+143 b (7 b+6)+42 f^2 \Big)\Bigg]
        \notag\\[5pt]
        &&\hspace{-1cm}~~
        -P_m(k_2) P_m(k_3) k_1 \Bigg[k_1 k_2 \bigg(k_1 k_2 \cos 2 \theta_{12} \Big(65 b (7 f+11)+3 f (35 f+13)\Big) +7 \cos \theta_{12} \Big(k_1^2 \big(143 b (f+1)
        \notag\\[5pt]
        &&\hspace{-1cm}~~
        +f (45 f+13))-286 b_2 k_2^2\Big)\bigg) +k_2^2 k_1^2 \Big(26 b (21 f+11)-1001 b_2+2 f (105 f+26)\Big) +21 k_1^4 f (13 b+3 f) -1001 b_2 k_2^4 \Bigg]
        \notag\\[5pt]
        &&\hspace{-1cm}~~
        - P_m(k_3) P_m(k_1) k_2^4  \Bigg[k_1 \Big(39 (11 b+f) \cos 2 \theta_{12}+26 (7 b+2) f+143 b (4-7 b)+63 f^2\Big)
        \notag\\[5pt]
        &&\hspace{-1cm}~~
        +7 k_2 \cos \theta_{12} \Big(13 b (5 f+11)+f (15 f+13)\Big) \Bigg]   \Bigg\}
  \end{eqnarray}
}

\end{document}